\documentclass[superscriptaddress,floatfix,aps,pre,twocolumn,notitlepage,shortbibliography,nofootinbib]{revtex4-2}

\usepackage[utf8]{inputenc}
\usepackage{amsmath}
\usepackage{amssymb}
\usepackage{refcount}
\usepackage{siunitx}
\usepackage{graphicx}
\usepackage{hyperref}
\hypersetup{colorlinks,allcolors=black}

\usepackage{natbib}
 \usepackage[capitalise]{cleveref}
\usepackage{microtype}
\usepackage{bm}
\usepackage{lipsum}
\usepackage[english]{babel}
\usepackage{color}
\usepackage{graphicx}
\usepackage{xcolor}
\usepackage{braket}

\begin{document}
\author{Haidar Al-Naseri}
\email{hnaseri@stanford.edu}
\affiliation{Stanford PULSE Institute, SLAC National Accelerator Laboratory, Menlo Park, California 94025, USA}

\title{Anisotropic momentum distributions due to radiation recoil in relativistic plasmas with electric and magnetic fields}
\pacs{52.25.Dg, 52.27.Ny, 52.25.Xz, 03.50.De, 03.65.Sq, 03.30.+p}

\begin{abstract}
The interaction of strong electromagnetic fields with plasma generates radiation accompanied by a recoil force, which can significantly alter the plasma dynamics. In this work, we investigate the development of anisotropic momentum distributions induced by the combined action of electric and magnetic fields on a thermally relativistic plasma. We consider three distinct types of anisotropy.
The first arises from a pure magnetic field acting on plasma with either isotropic or anisotropic initial momentum distributions, producing the characteristic ring-shaped momentum profile. The second is driven by a pure electric field, where radiation reaction generates a partial, anisotropic ring distribution in momentum space: significant modifications occur primarily in the 90°–180° and 270°–360° sectors of the $p_x-p_y$-plane, while the remaining quadrants remain largely unaffected.
The third case considers the combined effect of electric and magnetic 
fields. When the cyclotron frequency is very close to the upper hybrid 
frequency, the azimuthal symmetry of the ring-momentum distribution is 
broken. Conversely, in the regime where the cyclotron frequency is lower than the upper hybrid frequency, the rapid oscillations of the 
electric field dominate and preserve the symmetry of the ring-momentum 
distribution.

\end{abstract}
 
\maketitle

\section{Introduction}
The study of the interplay between quantum electrodynamics (QED) and 
collective plasma dynamics has gained considerable momentum over the 
past decade~\cite{QEDPlasma1,QEDPlasma2,QEDPlasma3,RingMomentum1,RingMomentum2,bulanov2024energy,Fiuza2023,gong2019radiation,Lorenzo2021,Uzdensky2023,liseykina2016inverse,zhang2020relativistic}. 
This growing interest is largely driven by rapid advances in high-intensity 
laser technology~\cite{E320,SLAC1997,Rutherford,QED-review1,QED-review2,QED-review3}, 
which now enable experiments in regimes where both QED and plasma effects 
play a central role. In astrophysical settings, extreme environments near 
compact objects such as magnetars naturally give rise to the same physics, 
and similar conditions are expected to be reproduced in forthcoming 
next-generation strong-field laboratory experiments~\cite{Fiuza2023}.  

The interaction of electromagnetic fields with plasma drives particles into 
relativistic motion. Accelerating charged particles lead to photon emission, 
and depending on the photon energy, this radiation can substantially alter 
the dynamics of the emitting particle. The associated recoil force, known 
as radiation reaction, has been extensively studied in both classical and 
quantum frameworks. In the classical regime, radiation reaction is often 
described by the Abraham--Lorentz--Dirac (ALD) equation, though it is well 
known to exhibit unphysical runaway solutions \cite{Runaway}. When radiation reaction acts 
only as a small perturbation, this difficulty can be circumvented by adopting 
the Landau--Lifshitz (LL) equation~\cite{Landau_Lif}. Exact solutions of the 
LL equation have been obtained for single-particle motion in specific field 
configurations~\cite{Bulanov,piazza2008exact}, while in plasma physics, the LL equation has 
been widely implemented in Particle-In-Cell (PIC) simulations, both in its 
classical form~\cite{Silva} and in quantum-corrected generalizations~\cite{Wallin,Silva_Q}.  

The incorporation of radiation reaction into kinetic theory was first 
established by Hakim \textit{et al.} in the framework of many-particle 
systems~\cite{Hakim}. Subsequent studies developed alternative derivations 
of radiation reaction--corrected Vlasov equations~\cite{Burton,Kunze,Elskens}, 
with the primary focus on the formal construction of kinetic evolution 
equations rather than their plasma dynamical consequences. A kinetic 
analysis of Landau damping modified by radiation reaction was presented in 
Ref.~\cite{Burton-2}. In parallel, hydrodynamic models of relativistic plasmas 
that incorporate radiation reaction have also been proposed; see, for 
instance, Refs.~\cite{Mahajan1,Mahajan2,Mahajan3,Dalakishvili}. More 
recently, self-consistent treatments of radiation reaction in kinetic plasmas 
have been investigated: the electrostatic limit was examined in 
Ref.~\cite{2023RR}, while the case of a circularly polarized electric field was 
studied in Ref.~\cite{RR2025}. For a plasma subjected to a strong external magnetic field—conditions typical of astrophysical plasmas—electrons can lose a significant portion of their kinetic energy. Previous studies have shown that radiation reaction in a constant and homogeneous magnetic field can drive a plasma initially in equilibrium to develop anisotropy, leading to the formation of a ring-shaped momentum distribution \cite{RingMomentum1, RingMomentum2}. However, the emergence of such ring-shaped anisotropy has not yet been explored in the case of radiation driven by an electric field, nor in situations where both electric and magnetic fields act simultaneously on the plasma.

In this work, we address this gap by investigating radiation reaction 
effects in plasmas subject to self-consistent electric fields in combination 
with external magnetic fields, with the aim of understanding the resulting 
anisotropy formation and momentum-space dynamics. We solve the relativistic 
Vlasov equation coupled with Ampère's law, allowing the plasma current to 
self-consistently modify the electric field. Our objective is to study how 
radiation reaction influences the momentum distribution for both initially 
isotropic and anisotropic plasmas.  
The radiation reaction force is modeled using the Landau--Lifshitz (LL) 
equation. The magnetic field is treated as constant and homogeneous, 
generated by an external source; thus, only the plasma contribution to the 
electric field is considered, while its contribution to the magnetic field is 
neglected. This treatment of the magnetic field is consistent with previous 
works \cite{RingMomentum1,RingMomentum2}, but our formalism additionally 
incorporates the effect of the self-consistent electric field as well as the 
magnetic force term $\mathbf{v}\times \mathbf{B}$, where $\mathbf{v}$ is the 
velocity of the electrons in the plasma and $\mathbf{B}$ is the magnetic field.  
In the case of pure synchrotron radiation with an initially isotropic plasma, 
it is often convenient to neglect the magnetic force term, as it does not 
affect the plasma dynamics. However, for an initially anisotropic plasma 
subject to a strong magnetic field, or in the combined presence of electric 
and magnetic fields, the magnetic force term must be retained.

The radiation reaction force is a non-conservative force, i.e., it does not 
conserve phase-space volume, unlike the Lorentz force. Depending on the 
geometry of the electromagnetic field, the development of anisotropy in 
the plasma momentum distribution can take different forms. By considering 
the combined effects of electric and magnetic fields in plasma, we are able 
to study distinct types of anisotropic structures.  
In this work, we investigate three different cases. First, we study pure 
synchrotron radiation for both isotropic and anisotropic initial plasmas. 
We find that an initial anisotropy in the plasma momentum distribution 
reduces the overall energy loss due to radiation, leading to a smaller 
ring-momentum distribution compared to the isotropic case.  
In the second case, we examine how radiation induced by an electric field 
gives rise to a new form of anisotropy, distinct from that of synchrotron 
radiation. This anisotropy exhibits similarities to the magnetic-field 
case in certain regions of momentum space, while in other regions the 
initial momentum distribution remains largely unaffected. Interestingly, 
the characteristic size of the anisotropic structure in momentum space is 
found to be comparable in both cases, provided that the initial electric 
and magnetic fields have the same energy.  
In the third case, we examine the combined influence of electric and magnetic fields on radiation reaction and its impact on the plasma momentum distribution. When the cyclotron 
frequency is close to resonance with the upper hybrid frequency, the 
ring-momentum structure associated with pure synchrotron radiation can 
be strongly modified. In contrast, if the cyclotron frequency is much 
smaller than the upper hybrid frequency, the modification of the 
ring-momentum structure remains weak.

The organization of this paper is as follows. In \cref{Theory}, we present 
the kinetic equation including the radiation reaction force considered in 
this work. In \cref{NumericalSection}, we present numerical simulation 
results for the three different cases of anisotropy development. Finally, 
in \cref{DiscussionSection}, we discuss the main findings of the study and 
provide concluding remarks.

\section{Kinetic theory including a radiation reaction force}
\label{Theory}
For sufficiently strong electromagnetic fields, the relativistic Vlasov equation 
describing the dynamics of an electron ensemble must be modified to incorporate 
quantum effects (see, e.g., Refs.~\cite{QED-review1,QED-review2,QED-review3,E-schwinger,al2021plasma}). 
When the field strength is well below the Schwinger critical value, electron--positron 
pair production via the Schwinger mechanism can be neglected. Nevertheless, photon 
emission by individual electrons through nonlinear Compton scattering may become 
significant whenever the parameter $\chi A_0^2$ is not too small. Here, we define 
the quantum parameter

\begin{equation}
    \chi=\frac{\gamma}{E_{cr}}\sqrt{|\mathbf{E}+ \frac{\mathbf{v}}{c}\times \mathbf{B}|^2- \Big(\frac{\mathbf{v}}{c}\cdot \mathbf{E} \Big)}
\end{equation}
Here, $\gamma$ denotes the Lorentz factor, and 
$E_{\mathrm{cr}} = m^{2} c^{3} / (|e| \hbar)$ is the critical field, 
where $e$ and $m$ are the electron charge and mass, respectively, 
$\hbar$ is the reduced Planck constant, and $c$ is the speed of light 
in vacuum. We also define the dimensionless laser strength parameter as 
$A_{0} = eE / (m \omega)$, with $\omega$ denoting the wave frequency.  
Nonlinear Compton scattering has been extensively investigated in the low-density plasma regime, where the electron density is sufficiently small such that the electromagnetic fields can be treated as vacuum solutions of Maxwell’s equations. In this limit, attention has largely been directed toward analyzing the characteristics of the emitted radiation spectra (see the recent reviews~\cite{QED-review1,QED-review2,QED-review3} for comprehensive references).  
At higher electron densities, however, plasma dynamics in strong fields becomes considerably more intricate, as self-consistent plasma currents must be included. Furthermore, if the radiated energy is significant, the electron dynamics---ordinarily governed by the Lorentz force---must be corrected to account for the radiation reaction force~\cite{Wallin,Silva,Silva_Q}.  

In this work, we model the plasma dynamics using the relativistic Vlasov equation, modified by an additional term representing radiation reaction.
Before presenting the kinetic equation, we define the normalized notation
\begin{multline}
    t_n=\omega_{ce} t, p_n=\frac{p}{mc}, \epsilon_n=\frac{\epsilon}{mc^2}\\
    f_n=\frac{m^3c^3}{n_0}f, E_n=\frac{eE}{mc\omega_{ce}}, B_n=\frac{eB}{m\omega_{ce}}, n_{0n}=\frac{c^3}{\omega_{ce}^3}n_0
\end{multline}
where $\omega_{ce}$ is the Compton frequency, $n_0$ is the plasma density, $\epsilon=\sqrt{1+p^2}$ and $f$ is the distribution of the plasma in the momentum space.
 Note that we will drop the subscript
n denoting the normalized quantities in what follows to
simplify the notation.
In this work, the radiation reaction force $\mathbf{F}_R$ is added to the Lorentz Force $\mathbf{F}_L=\mathbf{E}+ \frac{\mathbf{p}}{\epsilon}\times \mathbf{B}$ in the relativistic Vlasov equation as
\begin{equation}
\label{Kinetic_equation}
   \bigg[ \frac{\partial }{\partial t}+\frac{\mathbf{p}}{\epsilon}\cdot \nabla_r\bigg]f+
  \nabla_p \cdot \bigg((\mathbf{F_L}+\mathbf{F}_R)f\bigg)=0
\end{equation}

where  the radiation reaction force is given by
\begin{multline}
    \mathbf{F}_R=\frac{2\alpha}{3}\Bigg[\epsilon \frac{d\mathbf{E}}{dt}+\bigg(\frac{\mathbf{p}\cdot \mathbf{E}}{\epsilon}\bigg)\mathbf{E}+\mathbf{E}\times \mathbf{B}+\mathbf{B}\times \bigg(\mathbf{B}\times \frac{\mathbf{p}}{\epsilon}\bigg)\\
    -\epsilon \mathbf{p}\bigg(
    \Big(\mathbf{E}+\frac{\mathbf{p}}{\epsilon}\times \mathbf{B}\Big)^2
    -\Big(\frac{\mathbf{p}}{\epsilon}\cdot\mathbf{E}\Big)^2
    \bigg)
    \Bigg]
\end{multline}
where $\alpha$ is the fine-structure constant.
From this point onward, we focus on a field configuration where, in a 
suitable reference frame, the electromagnetic field is purely electric in 
nature. In addition, we assume a constant and homogeneous magnetic field. 
Such a configuration is representative of fields present in certain 
astrophysical environments, where the timescale governing the magnetic 
field variation is much longer than that of the electric field. The latter can be associated with coherent photon emission from plasma surrounding 
dense objects, such as magnetars \cite{uzdensky2014plasma}. Without loss of generality, we set $\mathbf{B} = B_0 \mathbf{e}_z$. The contribution of the plasma current to the magnetic field is neglected, and the magnetic field 
is assumed to be generated by an external source. In this case, Faraday’s 
law can be disregarded, and the system is closed using Ampère’s law with a vanishing time-dependent magnetic field; that is, we use

\begin{equation}
\label{Ampers1}
    \partial_t{\bf E}=-4\pi \alpha n_0 \int d^3p \frac{{\bf p}}{\epsilon}f
\end{equation}
We consider electric field that is perpendicular to the magnetic field, i.e. $\mathbf{E}=E_x\mathbf{e}_x+E_y\mathbf{e}_y$.
With these assumptions, the kinetic equation
in \cref{Kinetic_equation} is reduced to
\begin{multline}
\label{Kinetic_equation2}
    \frac{\partial f}{\partial t}+ 
    E_x\frac{\partial f}{\partial p_x}+ E_y\frac{\partial f}{\partial p_y}
   + \frac{B_0}{\epsilon} \bigg[p_y\frac{\partial f}{\partial p_x}-p_x\frac{\partial f}{\partial p_y} \bigg]\\= -\mathbf{F}_R\cdot \nabla_pf-f\nabla_p\cdot \mathbf{F}_R
\end{multline}
where the radiation reaction force is 

\begin{multline}
\label{RadiationForce_x}
      F_{R,x}=\frac{2\alpha}{3}\Bigg[\epsilon \frac{dE_x }{dt}+\frac{E_x}{\epsilon}\Big(p_xE_x+p_yE_y\Big)+B_0E_y-B_0^2\frac{p_x}{\epsilon}\\
    -\epsilon p_x\bigg(
    \Big(E_x-B_0\frac{p_y}{\epsilon}\Big)^2+ \Big(E_y+B_0\frac{p_x}{\epsilon}\Big)^2\\
    -\frac{1}{\epsilon^2}\Big(p_xE_x+p_yE_y\Big)^2
    \bigg)
    \Bigg]
\end{multline}
\begin{multline}
\label{RadiationForce_y}
      F_{R,y}=\frac{2\alpha}{3}\Bigg[\epsilon \frac{dE_y }{dt}+\frac{E_y}{\epsilon}\Big(p_xE_x+p_yE_y\Big)-B_0E_x-B_0^2\frac{p_y}{\epsilon}\\
    -\epsilon p_y\bigg(
    \Big(E_x-B_0\frac{p_y}{\epsilon}\Big)^2+ \Big(E_y+B_0\frac{p_x}{\epsilon}\Big)^2\\
    -\frac{1}{\epsilon^2}\Big(p_xE_x+p_yE_y\Big)^2
    \bigg)
    \Bigg]
\end{multline}

Before studying the case of electric and magnetic field in plasma, 
 we will look for an analytical solution of \cref{Kinetic_equation2} in the limit of vanishing $E$-field, in which the kinetic equation including radiation reaction \cref{Kinetic_equation2} reduces to
\begin{equation}
\label{Kinetic_equation3}
    \frac{\partial f}{\partial t}
  = -\mathbf{F}_R\cdot \nabla_pf-f\nabla_p\cdot \mathbf{F}_R
\end{equation}
where the radiation reaction term is $  \mathbf{F}_R=-2\alpha B_0^2 \mathbf{p} (1+p_{\perp}^2)/3\epsilon$, here $p_{\perp}$ is the perpendicular momentum. Expressing $\nabla_p$ in cylindrical momentum coordinates, and assuming that the initial plasma distribution is isotropic, i.e. the magnetic force term vanishes, the kinetic equation reduces to
\begin{equation}
\label{Diff_equation}
    \frac{3}{2 \alpha B_0^2}\frac{\partial f}{\partial t}=\eta \frac{p_{\perp}^2}{\epsilon}f+ \frac{p_{\perp}^2}{\epsilon}\frac{\partial f}{\partial p_{\perp}}+\frac{p_{\perp}^2p_z}{\epsilon}\frac{\partial f}{\partial p_z}
\end{equation}
where $\eta=5-(p_{\perp}^2+p_{z}^2)/\epsilon^2$. In the ultra-relativistic limit, the factor $\eta=4$. As shown in \cite{RingMomentum1,RingMomentum2}, using the method of characteristic, the analytical solution to \cref{Diff_equation} is for any arbitrary initial distribution $f_0$ is
\begin{equation}
\label{AnalyticalResultPlasma}
    f(p_{\perp},p_z,t)=\frac{f_0\Big(\frac{p_{\perp}}{1-\tau p_{\perp}},\frac{p_z}{1-\tau p_{\perp}}\Big)}{(1-\tau p_{\perp})^4}
\end{equation}
where $\tau= \frac{2}{3}\alpha B_0^2 t$. This equation fully determines the time dependence of the distribution 
function undergoing synchrotron radiation. The solution is characterized 
by an upper bound $\tilde{p}_{\perp}$, defined as $\tilde{p}_{\perp} = \tau^{-1}$, 
such that $p_{\perp}\tau < 1$. In other words, the distribution function 
is progressively compressed, with the upper bound $\tilde{p}_{\perp}$ 
decreasing over time.
As studied in \cite{RingMomentum1,RingMomentum2}, an initially isotropic relativistic thermal plasma tends to develop anisotropy due to radiation reaction. 
The time evolution of the plasma distribution, presented in 
\cref{AnalyticalResultPlasma}, exhibits an inverted Landau population; 
that is, there exist regions in the perpendicular momentum $p_{\perp}$ 
where $\partial f / \partial p_{\perp} > 0$. This behavior leads to the 
formation of an anisotropy in momentum space, characterized by a 
ring-shaped momentum distribution. We define the radius of the ring, $p_R$, 
by the condition
\[
\left.\frac{\partial f(p_{\perp},t)}{\partial p_{\perp}}\right|_{p_{\perp}=p_R} = 0,
\]
and, using \cref{AnalyticalResultPlasma}, we obtain

\begin{equation}
\label{RingRadiusEquation}
    p_{R} = \frac{1 + 6 p_{\mathrm{th}}^2 \tau^2 - \sqrt{1 + 12 p_{\mathrm{th}}^2 \tau^2}}{6 p_{\mathrm{th}}^2 \tau^3}.
\end{equation}
This equation describes the time-dependence of the ring momentum radius developed due to pure synchrotron radiation.
The ring radius is initially zero and increases over time as a result of 
the cooling mechanism. It reaches a maximum at the time determined by the 
condition $\left.\partial p_R(t)/\partial t\right|_{t=t_R} = 0$, where 
\begin{equation}
    t_R = \frac{3}{4 \alpha B_0^2 p_{th}} ,
\end{equation}
denotes the characteristic time required to attain the maximum ring radius 
in the pure synchrotron radiation. However, $t_R$ can still be used to describe the characteristic timescale of the anisotropic development of the electric field, in which case the amplitude of $B_0$ is simply replaced by $E_0$. Later in this paper, we will employ $t_R$ in the combined case of electric and magnetic fields, where only the magnetic field amplitude is considered.
Beyond the maximum value of the ring radius, continued radiation causes the ring radius to decrease.
Later in this work, we extend the conclusions from 
\cref{AnalyticalResultPlasma} to investigate the physics of the inverted 
Landau population in three distinct cases. The first case considers a 
constant external magnetic field with the plasma initialized in both 
isotropic and anisotropic configurations. The second case examines a 
relativistic thermal plasma subject to a time-varying electric field in the 
absence of a magnetic field. Finally, the third case addresses the 
situation where both electric and magnetic fields are present 
simultaneously.

\section{Numerical simulations}
\label{NumericalSection}
\subsection{Preliminaries}
The kinetic system consisting of the Vlasov equation in \cref{Kinetic_equation2} 
and Ampère’s law in \cref{Ampers1} is solved numerically in two-dimensional momentum 
space with coordinates $p_x$ and $p_y$. We assume that the parallel momentum component $p_z$ is zero in order 
to simplify the calculation, without losing any relevant physics, since 
the system does not exhibit parallel acceleration.
Spatial dependence is neglected as well, because there is no 
time-varying magnetic field—only a constant magnetic field 
$\mathbf{B} = B_0 \mathbf{e}_z$ and a time-varying electric field.  
The time evolution of the distribution function $f(t,\mathbf{p})$, and hence of 
the self-consistent electric field (i.e., the system \cref{Kinetic_equation2,Ampers1}), 
is computed in two steps. First, the Vlasov advection due to the Lorentz force 
is solved using the method of characteristics. The distribution function is interpolated 
from its previous value at each momentum grid point $(p_x,p_y)$ according to  
\[
f^{*}(p_x,p_y,t+\Delta t) = 
f\!\left(p_x - e F_x \Delta t, \, p_y - e F_y \Delta t, \, t\right),
\]  
where $F_x$ and $F_y$ denote the Lorentz force components in the $x$- and $y$-directions. 
This procedure ensures phase-space density conservation along particle trajectories.  

In the second step, radiation reaction is included through the conservation equation  
\[
\frac{\partial f}{\partial t} + \nabla_{\mathbf{p}} \cdot \left( \mathbf{F}_{R} f \right) = 0,
\]  
where the interpolated distribution $f^{*}$ is used as the initial condition. 
This equation is solved numerically using a fourth-order Runge–Kutta (RK4) method, 
with flux divergences evaluated by finite differences on the momentum grid.  
Finally, a stabilization step is applied to remove small negative values of $f$ 
(arising from interpolation) and to truncate the high-momentum tail in order to 
prevent numerical instabilities. The resulting distribution $f$ 
therefore incorporates both Vlasov advection and radiation reaction effects 
at each timestep.

Through the paper, we consider a plasma that is initially in thermal equilibrium and follows a Maxwell–Boltzmann distribution.
\begin{equation}
\label{Maxwellian_distribution}
    f_0= c_n e^{-p_x^2/p_{th,x}^2}e^{-p_y^2/p_{th,y}^2}
\end{equation}
where $p_{\text{th},x}$ and $p_{\text{th},y}$ denote the thermal spread in the $x$- and $y$-directions, respectively, and $c$ is the normalization constant, given by $c_n = 1/\int dp_x dp_y e^{-p_x^2/p_{th,x}^2}e^{-p_y^2/p_{th,y}^2}$.

\subsection{Pure synchrotron radiation}
\label{SynchrotronRadiation}
We consider here the case of pure synchrotron radiation; that is, we assume the absence of any electric field and the presence of only a constant magnetic field. Our aim is twofold: first, to benchmark our results against the analytical solution in \cref{AnalyticalResultPlasma} and the numerical results reported in \cite{RingMomentum1,RingMomentum2}; and second, to investigate how an initially anisotropic plasma evolves under radiation and whether a similar inverted Landau population emerges.
The radiated power scales as $P \propto \gamma^{4} (\mathbf{p} \times \mathbf{a})^{2}$, where $\mathbf{a}$ is the acceleration of the electron. Since the system contains no electric field, the plasma must be relativistically hot, i.e., $p_{\mathrm{th}} \gg 1$, in order to produce significant radiation.
The relevant quantum parameter in this configuration is given by $\chi = B_{0} p_{\perp}$. 

Considering an isotropic plasma with $p_{\text{th},x} = p_{\text{th},y} = 20$ and $B_0 = 0.05$, the quantum parameter is $\chi \leq 1$ for most particles in the plasma. Although the Landau–Lifshitz (LL) equation is derived under the classical assumption $\chi \ll 1$, it has been shown in \cite{RingMomentum1,RingMomentum2} that, for synchrotron radiation, the LL model agrees well with the full quantum radiation model even at $\chi \simeq 1$ as the correction due to quantum radiation is small compared to the classical one. While some discrepancies start to arise when entering the fully quantum regime, the energy loss due to radiation reduces the thermal momentum of the particles reducing the quantum parameter $\chi$ and the radiation can be described by the classical model.
Furthermore, as demonstrated in \cite{RR2025}, the agreement between Landau-Lifshitz and quantum radiation improves for higher initial plasma temperatures, that is for relativistic hot plasma such as the ones considered in this work, the Landau-Lifshitz model should be valid at $\chi \sim 1$.

\begin{figure}
    \centering
    \includegraphics[width=8.5 cm, height=10 cm]{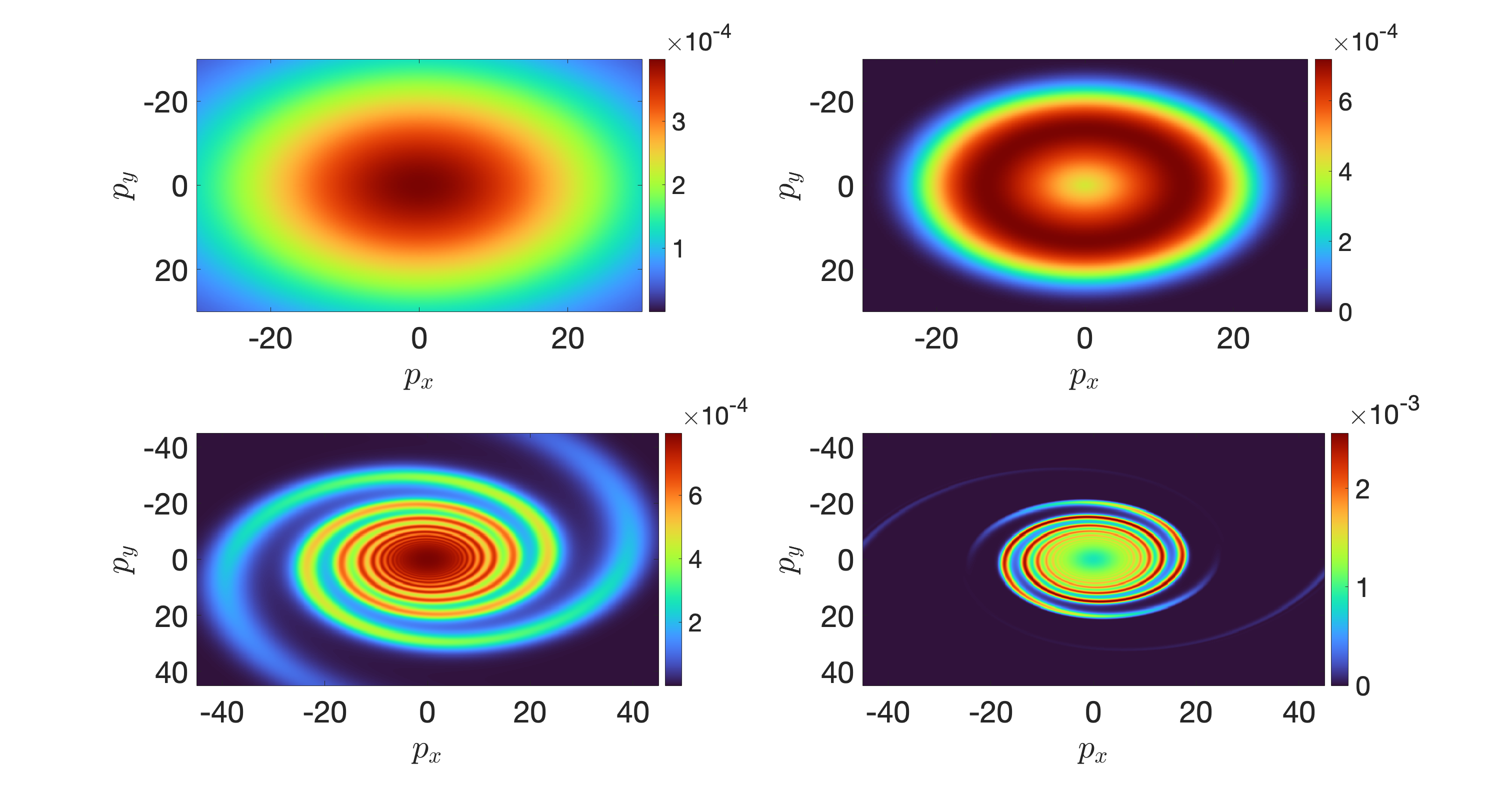}
    \caption{Momentum distribution of the plasma. In the upper row, we used $p_{th,x}=p_{th,y}=20$, where in the left figure we display the plasma at $t=0$ and in the right show the distribution at $t=t_R$. In the lower row, we used $p_{th,x}=20$ and $p_{th,y}=10$. For the left figure, we set the radiation reaction force $\mathbf{F}_R=0$ to display the magnetic force effect, while in the right figure, we included the magnetic and radiation force. For all figures, we used $B_0=0.05$.}
    \label{Fig1}
\end{figure}

In the first row of \cref{Fig1}, we present the momentum distribution of the isotropic plasma. The left panel shows the initial distribution, while the right panel corresponds to the distribution at $t = t_R$, when the radius of the ring momentum reaches its maximum. Comparing the two panels, we observe the formation of a ring structure in the right panel. The radius of the ring is approximately 13.3, which agrees well with the corresponding value in the analytical result in \cref{RingRadiusEquation}. 
Radiative cooling is clearly visible in the right panel, as the distribution becomes concentrated at lower momenta with a higher population density (as indicated by the colorbar maximum), compared to the initial distribution. The inverted plasma distribution arises because electrons with higher perpendicular momentum lose energy more rapidly, as the radiation reaction force is stronger at higher $p_{\perp}$. As a result, these high-$p_{\perp}$ electrons decelerate and begin to bunch with lower-$p_{\perp}$ electrons, leading to an accumulation of particles at lower perpendicular momenta. As the radiation process continues, successive layers of higher-energy electrons slow down and bunch in the same region, overlapping with the earlier bunched particles. This progressive bunching into a narrower zone causes the peak of the distribution to shift toward higher $p_{\perp}$, not because the electrons are gaining energy, but because the location of maximum particle density moves outward in momentum space due to the collective convergence. Throughout this process, the total energy of the particles continues to decrease due to radiation losses.

In the lower row of \cref{Fig1}, we show the momentum distribution of a non-isotropic plasma with thermal momenta $p_{\text{th},x} = 20$ and $p_{\text{th},y} = 10$. Both the left and right panels correspond to the distribution at $t = t_R$.
In the left panel, the radiation force $F_R$ is set to zero in order to only display the effect of the magnetic force arising from the anisotropic thermal distribution. As seen, the plasma particles undergo gyromotion, resulting in a spiral-like structure in momentum space. The non-zero $\mathbf{v} \times \mathbf{B}$ force causes particles to drift from the $p_x$ to the $p_y$ direction.
In the right panel, we present the momentum distribution including the effects of radiative cooling. The spiral structure persists, but the distribution becomes more concentrated at lower momenta. In particular, the spirals at high momenta become narrower due to the decreased particle occupation at high momentum caused by radiation losses.
At lower momenta, we also observe the formation of a ring-like structure, similar to the one seen in the isotropic case but with a lower momentum radius $p_R=8.8$. This value is lower than for the case of isotropic plasma with $p_{th}=20$ but higher than the momentum radius of isotropic plasma $p_{th}=10$. One of the reasons why the ring radius is smaller in the non-isotropic case is the reduced energy loss due to radiation, compared to the isotropic case. Another contributing factor is the magnetic force, which induces momentum flux from the $p_x$ to the $p_y$ direction. This redistribution affects the electron bunching structure by modifying how particles accumulate in momentum space under the influence of radiation.

\begin{figure}
    \centering
    \includegraphics[width=8.5 cm, height=10 cm]{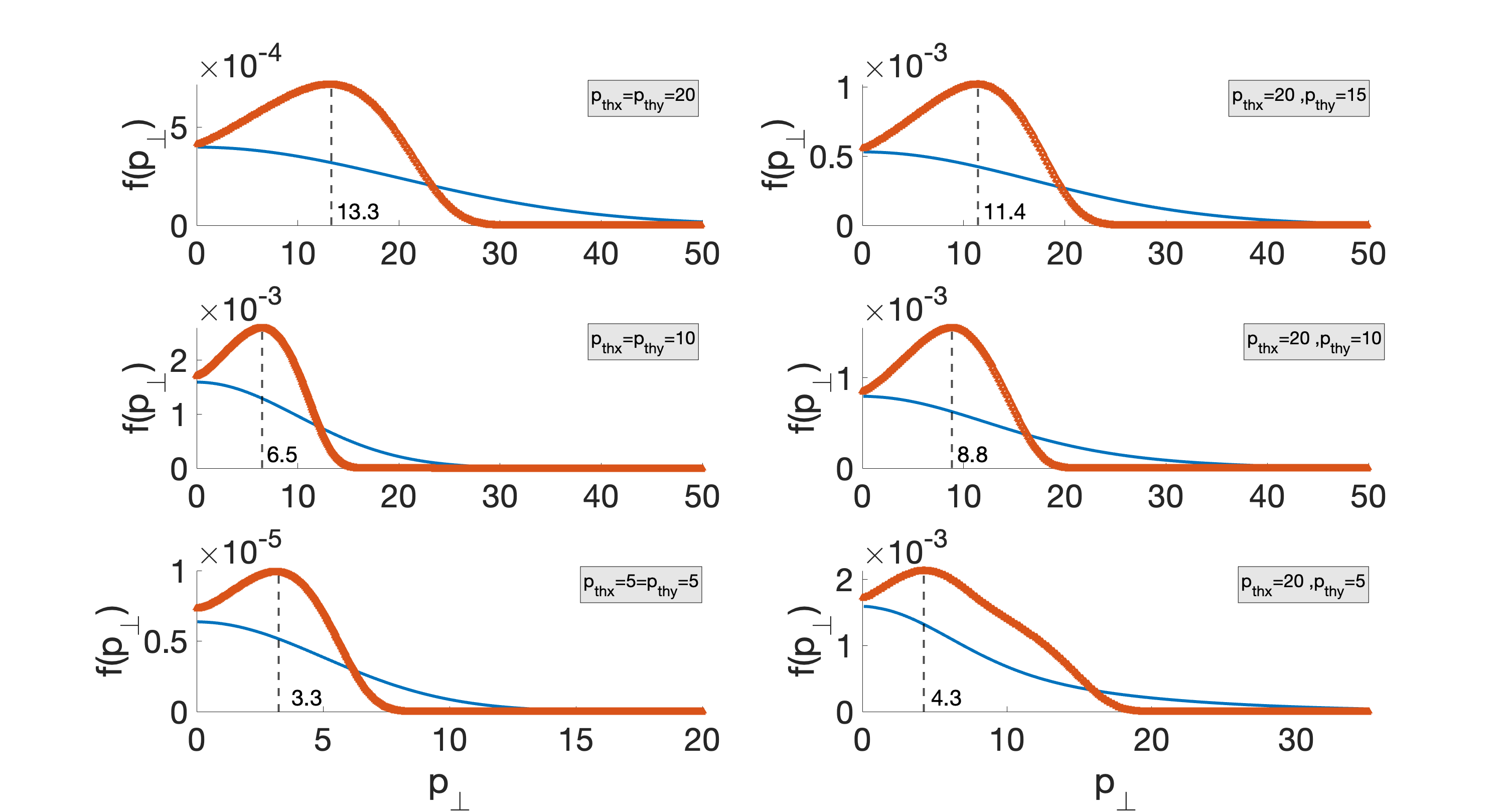}
    \caption{Perpendicular momentum distribution of the plasma, initial plasma (solid curve) and the distribution formed at the peak of the ring momentum formation (jagged curve). In the left column, we have isotropic initial distribution, while in the right column we have anisotropic distribution.  }
    \label{Fig2}
\end{figure}

To further investigate how anisotropy affects the plasma distribution under radiation, we display in \cref{Fig2} the dependence of different plasma distributions on the perpendicular momentum. In the left column of \cref{Fig2}, we show an initially isotropic plasma: both the initial distribution (solid curve) and the one formed at \(t = t_R\) (jagged curve). As expected, particles with initially high energies are cooled to lower energies due to radiation. However, the peak of the distribution does not shift to the lowest energies, as is typical for radiative cooling in plasmas. Instead, due to the development of a population inversion, higher energy states become more populated than the lowest ones.
In the right column of \cref{Fig2}, we present the initial and evolved plasma distributions for three anisotropic cases. In the upper panel, we set \(p_{\text{th},x} = 20\), \(p_{\text{th},y} = 15\),
in the middle panel \(p_{\text{th},x} = 20\), \(p_{\text{th},y} = 10\)
and in the lower panel, \(p_{\text{th},x} = 20\), \(p_{\text{th},y} = 5\). Due to the temperature imbalance between the \(x\)- and \(y\)-directions, the magnetic force induces a drift of particles in momentum space from the higher-temperature region to the lower-temperature one. This redistribution reduces the overall energy loss due to radiation, as evidenced by the higher particle occupancy at larger momentum values.
Comparing the high-momentum particle densities between the left and right columns, we observe a greater population at high momentum in the anisotropic cases. Furthermore, the distributions are less symmetric due to uneven energy loss, with certain regions of momentum space experiencing greater radiative damping than others, as a result of the gyromotion.
These effects reduce the radius of the ring-shaped momentum distribution in the anisotropic case compared to the isotropic one.

\begin{figure}
    \centering
    \includegraphics[width=8.5 cm, height=10 cm]{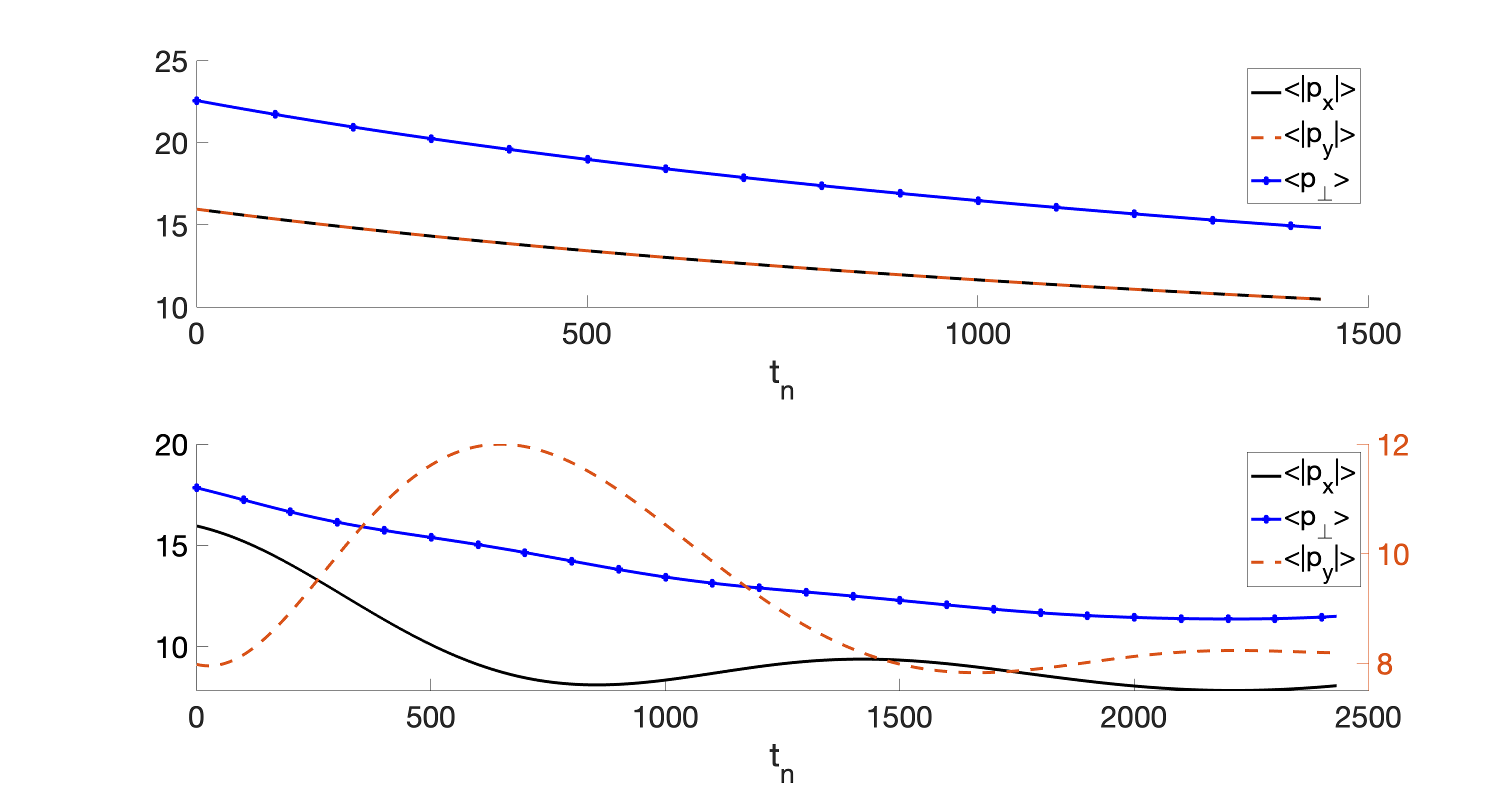}
\caption{
The average values of the momentum components $p_x$ (solid curve), $p_y$ 
(dashed curve), and the perpendicular momentum $p_{\perp}$ (curve with 
circles) are plotted as functions of time for two cases. The upper panel 
shows the isotropic case with $p_{\mathrm{th}} = 20$, while the lower panel 
presents the anisotropic case with $p_{\mathrm{th},x} = 20$ and 
$p_{\mathrm{th},y} = 10$.
}

    \label{Fig3}
\end{figure}

To understand how the drift of particles from high to low temperature affect the radiation, we study the average values of the change of $p_x$ and $p_y$ using
\begin{align}
    <|p_x|>&= \int dp_xdp_y |p_x| f(t)\\
     <|p_y|>&=\int dp_xdp_y |p_x| f(t)
\end{align}
For the perpendicular momentum, we use the definition \(\langle p_{\perp} \rangle = \sqrt{\langle p_x^2 \rangle + \langle p_y^2} \rangle\). In \cref{Fig3}, we plot all three quantities using two different initial distributions. In the upper panel, we use an isotropic distribution, \(p_{\text{th},x} = p_{\text{th},y} = 20\). As observed, the average values of $p_x$ and $p_y$, and consequently the 
average perpendicular momentum $p_{\perp}$, decrease over time. This 
dynamical behavior is a direct consequence of radiation reaction, which 
continuously reduces the particle momenta in both the $x$ and $y$ 
directions. As a result, the perpendicular momentum exhibits an almost 
constant rate of reduction.
In the lower panel of \cref{Fig3}, we consider an anisotropic initial 
plasma distribution with $p_{\text{th},x} = 20$ and $p_{\text{th},y} = 10$. Due to the unequal thermal spread in the $x$ and $y$ directions, the 
particle flow from $x$ to $y$ increases the average value of $p_x$ to 
approximately $1.5$ times its initial value. This redistribution reduces 
the rate of energy loss, as evidenced by the evolution of the average 
perpendicular momentum, which exhibits a smaller slope compared to the 
isotropic case.


\subsection{Anisotropic formation due to pure electric field}
In this subsection, we consider only an electric field propagating in plasma and assume that the magnetic field is zero. We use the same assumption of the Maxwell--Boltzmann distribution \cref{Maxwellian_distribution} as in the previous subsection. We let the plasma interact self-consistently with an electric field that initially has peak values in the $x$- and $y$-directions. As the electric field induces a current in the $x$- and $y$-directions, the contribution from the current updates the values of the electric field through Ampère's law \cref{Ampers1}. 

The radiation reaction force in the vanishing magnetic field-case is 
\begin{multline}
\label{RadiationForce_onlyE}
            F_{R,i}=\frac{2\alpha}{3}\Bigg[\epsilon \frac{dE_i }{dt}+\frac{E_i}{\epsilon}\Big(p_xE_x+p_yE_y\Big)\\
    -\epsilon p_i\bigg(
    E^2
    -\frac{1}{\epsilon^2}\Big(p_xE_x+p_yE_y\Big)^2
    \bigg)
    \Bigg]
\end{multline}
where $i=x,y$.
Starting with an electric field, it is possible to accelerate the plasma to ultra-relativistic energies and obtain significant radiation without starting with a relativistically hot plasma. However, to investigate whether a similar ring-formation physics exists in the case of an electric field in plasma, we have to start from a relativistically hot plasma. Furthermore, the Lorentz factor due to acceleration by the electric field must be much smaller than the thermal Lorentz factor, to ensure that the same condition applied to synchrotron radiation is now applied to radiation due to the electric field.

\begin{figure}
    \centering
    \includegraphics[width=8.5 cm, height=10 cm]{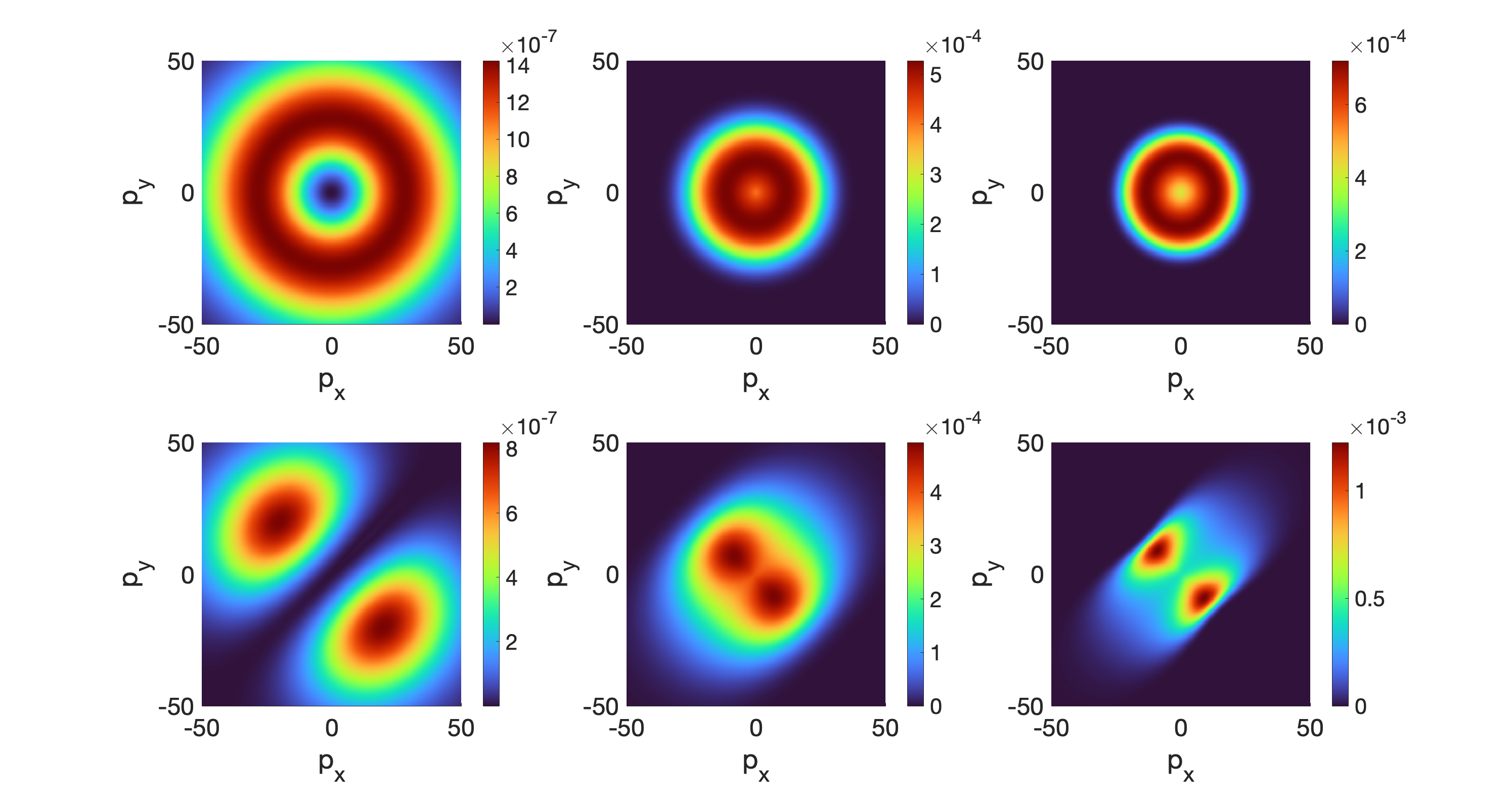}
    \caption{Demonstration of the evolution of the plasma distribution after radiation. In the first row, the figures are for the case of only magnetic field, while the lower panel is for only electric field.
    In the first column we display the magnitude of the radiation force times the initial plasma distribution $|\mathbf{F_{R}}|f_0$. In the second column we plot the plasma distribution at $t=0.4 t_R$ and in the third column we display the plasma distribution at $t=4 t_R$ }
    \label{Fig4}
\end{figure}

We perform a numerical simulation with $E_0 = 0.05$, the same magnitude as in the synchrotron radiation case, and start with an isotropic relativistically hot plasma with $p_{\text{th}} = 20$. To restrict the Lorentz factor due to the electric field, we had to run the simulation with a dense plasma. 
Starting with the plasma densities $n_0 = 7 \times 10^{28} \, \text{cm}^{-3}$ we obtained a Lorentz factor (not accounting for the thermal Lorentz factor) of the order of unity. 
In \cref{Fig4}, we show the results from the simulation of the plasma distribution after radiation due to the electric field, together with the results from the synchrotron radiation simulation. In the upper row of \cref{Fig4}, the plots correspond to the synchrotron simulation, while the lower row represents the plots from the electric field simulation. In the first column, we plot the magnitude of the radiation force multiplied by the initial plasma distribution, $|\mathbf{F}_R| f_0$. For the synchrotron radiation, it is clear that the radiation is azimuthally symmetric as we know that  $  \mathbf{F}_R=-2\alpha B_0^2 \mathbf{p} \epsilon/3$., with a peak at higher momenta of the plasma distribution, since the thermal Lorentz factor is larger there. In the lower row of the first column, the radiation force arising from the 
electric field is displayed. The radiation force is not azimuthally symmetric 
and exhibits pronounced peaks at large momenta, $p = \sqrt{p_x^2 + p_y^2}$. This behavior arises because the radiation force is dominated by the third term of \cref{RadiationForce_onlyE}, which is higher order in $p$ compared to the other two terms. We can rewrite the third term of 
\cref{RadiationForce_onlyE} as

\begin{equation}
\label{Radiation_force_strong_term}
    \mathbf{F}=-\frac{2\alpha \epsilon \mathbf{p}}{3}\bigg[E^2-\frac{(\mathbf{p}\cdot \mathbf{E})^2}{\epsilon^2}\bigg]
\end{equation}
Compared to the synchrotron case, when $E_0 = B_0$, the ratio of the third term in \cref{RadiationForce_onlyE} to the force term in the pure magnetic field case becomes
\begin{equation}
\label{Force_fraction}
    \frac{F_E}{F_B}=1-\frac{(p_xE_x+p_yE_y)^2 }{E_0^2\epsilon^2}
\end{equation}
The second term breaks the azimuthal symmetry observed in synchrotron radiation, 
since there are regions where $F_E/F_B$ can vanish. 
We assume $E_x(t=0)=E_y(t=0)=E_0$, which results in identical time evolution 
for $E_x$ and $E_y$. 
Because the plasma is relativistically thermal, we can approximate $p \sim \epsilon$. 
For particles in the distribution located at high $(p_x, p_y)$ and $(-p_x, -p_y)$, 
the second term in \cref{Force_fraction} cancels. 
In the remaining regions of momentum space, the second term in \cref{Force_fraction} 
is negligible, and the radiation reaction force behaves similarly to the pure magnetic field case.
This leads to the momentum distribution that we can see in the lower row of the first column of \cref{Fig4}

In the second column of \cref{Fig4}, we show how the plasma distribution evolves over time, at $t = 0.4\,t_R$. In the upper row, the ring momentum distribution is clearly visible, although its radius has not yet reached its maximum. The radiation force, which is strongest at higher momentum, pushes the particles toward lower momentum. As the particles bunch toward lower momentum, increased radiation leads to the overlapping of different particle layers, causing the peak momentum to shift toward higher energy.  
In the lower row, a similar process occurs, but with a different anisotropy. The azimuthally asymmetric radiation force pushes particles with higher perpendicular momentum toward lower momentum, where the radiation force is weaker. Over time, as more particles are pushed toward lower momentum, the peak momentum shifts to higher values, as in the synchrotron case, but with an asymmetric shape. In particular, particles located at high $(p_x, p_y)$ and high $(-p_x, -p_y)$ remain almost unaffected by radiation, compared to the synchrotron case where the radiation is azimuthally symmetric. This is because, at high $(\pm p_x, \pm p_y)$, the second term of \cref{Radiation_force_strong_term} nearly cancels the first term, resulting in minimal net radiation.

In the third column of \cref{Fig4}, we display the plasma distribution at $t = 4t_R$, where the plasma exhibits a new shape due to radiation effects. In the upper row, the radius of the ring momentum has already reached its maximum, and the higher-momentum states remain unoccupied. In the lower row, the plasma distribution resembles that of the synchrotron case when considering the azimuthal angles $90^\circ$--$180^\circ$ and $270^\circ$--$360^\circ$. At these angles, the momentum component $p_x$ has an opposite sign to $p_y$, which leads to a cancellation in the second term of \cref{Radiation_force_strong_term}, resulting in a strong radiation reaction force. Distinct peaks appear particularly at $135^\circ$ and $315^\circ$, whereas at $45^\circ$ and $225^\circ$, the plasma remains almost unchanged from its initial state, since the radiation force is minimal at these angles.

In the case of synchrotron radiation, the symmetry of the ring-shaped 
momentum distribution allows us to follow the time evolution of the 
anisotropic distribution by monitoring the evolution of its radius. For the case of an external electric field, a similar scaling of the anisotropy can be observed, but only when focusing on specific angles in the $p_x$--$p_y$ plane. As shown in the second and third panels of the lower row of \cref{Fig4}, the distance separating the two maxima increases with time, indicating that dynamics similar to those in the synchrotron case can also be studied here. 
 In the synchrotron case, we defined the radius of the ring-momentum as the value of the perpendicular momentum for which $\partial f / \partial p_{\perp} = 0$. In the case of electric field, this definition cannot be applied uniformly to all angles in the $p_x$--$p_y$ plane. However, by fixing the angle to $135^\circ$, we can define a radius-like distance in the same way as we did for the synchrotron case.

\begin{figure}
    \centering
    \includegraphics[width=8 cm, height=10 cm]{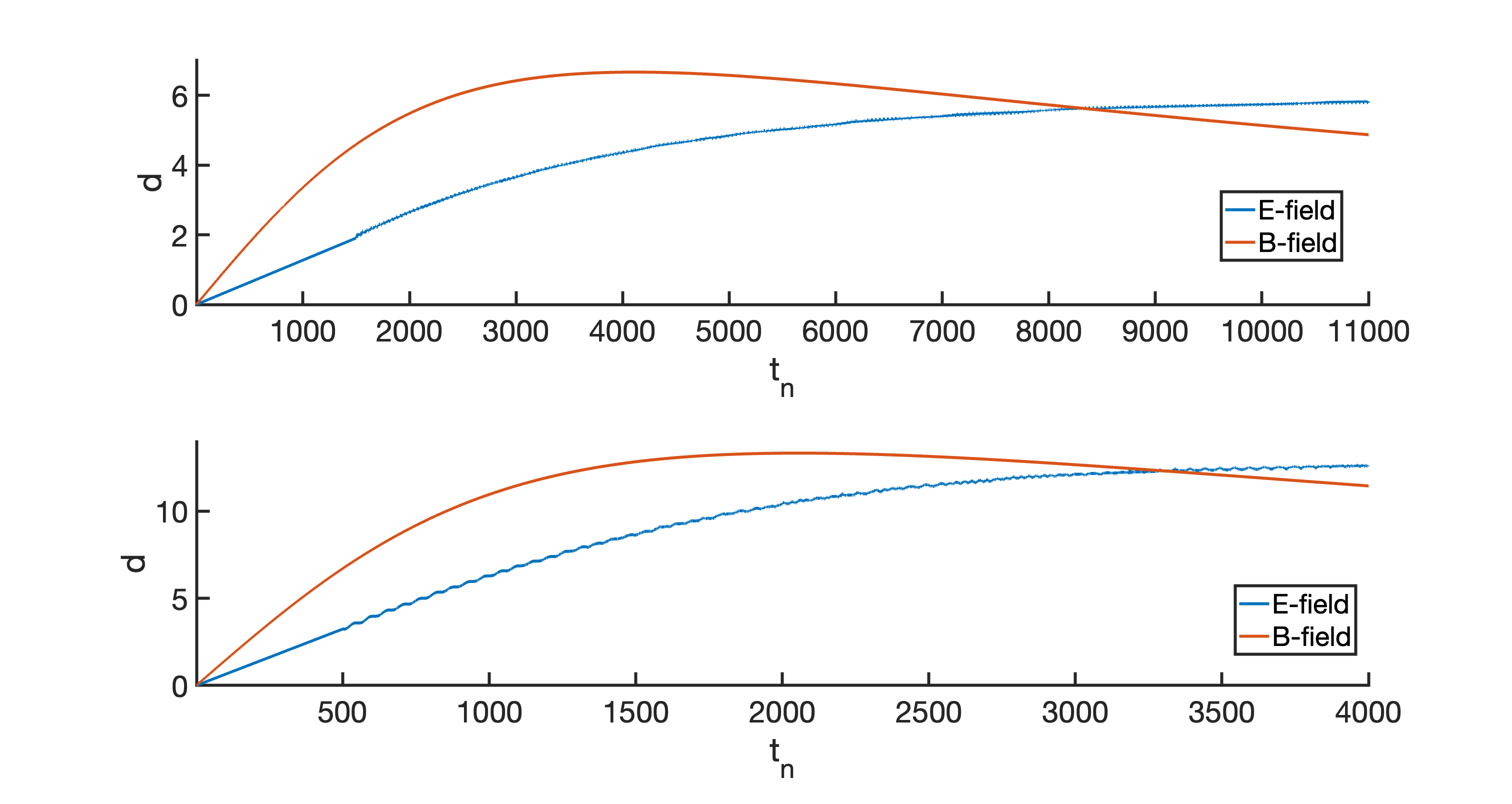}
    \caption{The distance $d$ for the ring momentum distribution for both cases of pure electric and magnetic field. In the first panel we used $p_{th}=10$ while in the second we had $p_{th}=20$. }
    \label{Fig5}
\end{figure}

In \cref{Fig5}, we show the time evolution of the ring-momentum anisotropy radius $d$ (denoted as $d$ rather than $r$, since in the electric-field case there is no symmetric structure) for the magnetic field case, together with the corresponding behavior in the electric-field case. In the upper panel, the initial plasma distribution follows a Maxwell–Boltzmann distribution with $p_{\mathrm{th}} = 10$. The radius in the magnetic-field case reaches a peak that is almost the same as in the electric-field case. However, the time required to reach the peak is longer in the electric-field case.
In the lower panel of \cref{Fig5}, we show the same information as in the upper panel, but for $p_{\mathrm{th}} = 20$. We observe nearly the same peak radius in both the electric- and magnetic-field cases, although the characteristic time scales differ.
The reason it takes longer for the electric field to reach a comparable peak is that the radiation process is slower. In the synchrotron case, the acceleration is constant in time, leading to a faster energy loss due to continuous radiation. In contrast, in the electric-field case the acceleration is maximized only when the field reaches its peak. Since the field oscillates in time, there are intervals when the field strength is below its maximum and the radiation is weaker. This results in a slower overall energy loss, and it takes longer for the bunching process to develop and produce the anisotropy.

\begin{figure}
    \centering
    \includegraphics[width=8.5 cm, height=10 cm]{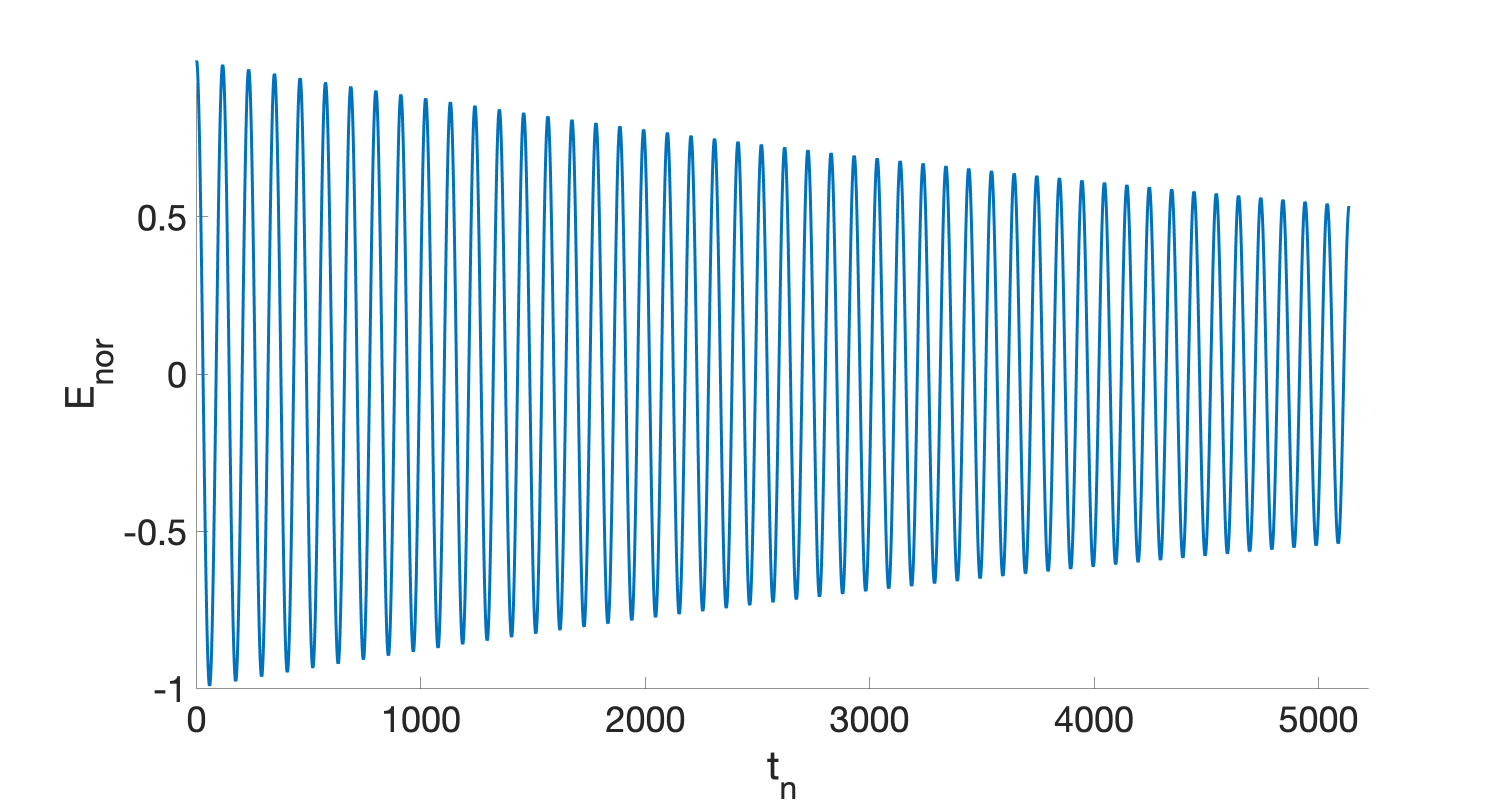}
    \caption{Time evolution of the normalized electric field, $E_{\mathrm{nor}}=E(t_n)/E_0$, 
as a function of time using the plasma density 
$n_0 = 7 \times 10^{28}\,\text{cm}^{-3}$.
 }
    \label{FigE}
\end{figure}

Before we end our analysis of the anisotropy formation in a pure electric field, we examine the damping of the field amplitude. Due to radiation, the energy stored in the electric field is damped over 
time. In \cref{FigE}, we show the time evolution of the electric field 
normalized to its initial value, using $p_{\mathrm{th}} = 20$, $E_0 = 0.05$, 
and $n_0 = 7 \times 10^{28}\,\text{cm}^{-3}$. As seen in the figure, the 
peak field decreases with time. In particular, at $t = 4000$---the time 
at which the radius of the ring momentum induced by radiation reaches its 
maximum---the field amplitude has decreased to $60\%$ of its initial value.

\subsection{Radiation due to electric and magnetic fields}

\begin{figure}
    \centering
    \includegraphics[width=8.5 cm, height=10 cm]{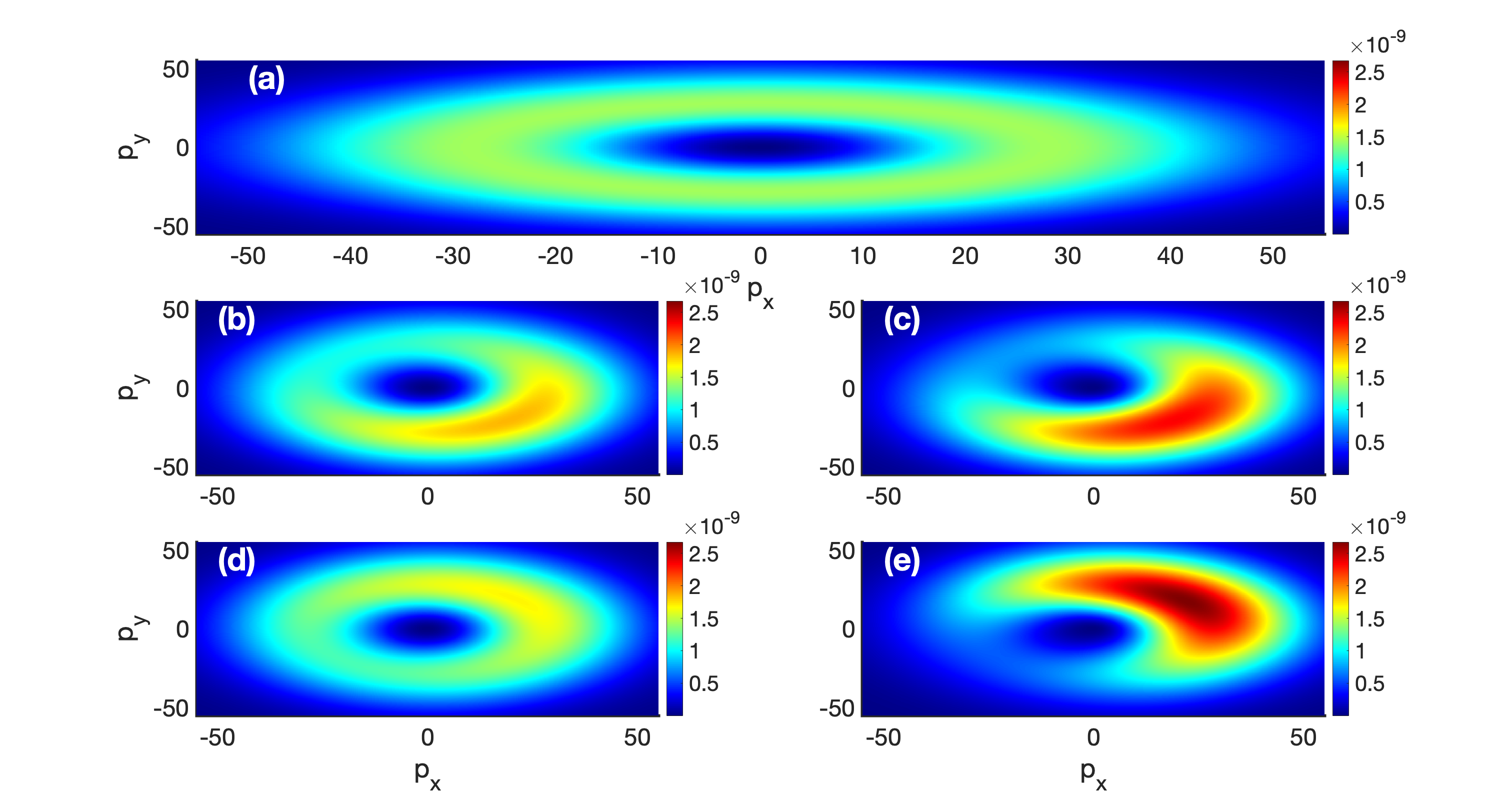}
    \caption{The momentum distribution of the radiation force in the plasma, given by $|\mathbf{F}_{\mathrm{R}}| f_0$, is shown in panels (a)–(e).  
In (a), we consider a pure magnetic field (i.e., no electric field).  
In (b), we use $E_0 = 0.005$, $B_0 = 0.05$, and $t = 0$.  
In (c), the parameters are $E_0 = 0.01$, $B_0 = 0.05$, and $t = 0$.  
In (d), we take $E_0 = 0.005$, $B_0 = 0.05$, and $t = 0.6 t_R$.  
Finally, in (e), we use $E_0 = 0.01$, $B_0 = 0.05$, and $t = 0.6 t_R$.  
 }
    \label{Fig6}
\end{figure}
In this subsection, we investigate the radiation effects on a plasma in the presence of both electric and magnetic fields.  
The electric field is treated self-consistently, as in the previous subsection, while the magnetic field is considered as an external and constant field, as discussed earlier in the manuscript.  
The initial plasma distribution is relativistic hot and is given by \cref{Maxwellian_distribution}.  
To analyze how the combination of the electric and magnetic fields affects radiation in a relativistic plasma, we start by examining the radiation force in \cref{RadiationForce_x,RadiationForce_y} in more detail.  
We consider a relativistic temperature with $p_{th} \gg 1$, and a dense plasma such that the gamma-factor $\gamma_E$ associated with the motion induced by the electric field satisfies $\gamma_E < \gamma_{th}$, where $\gamma_{th} = \sqrt{1+p_{th}^2}$.  
Thus, the dominant contribution in \cref{RadiationForce_x,RadiationForce_y} comes from the term of higher order in $p$ (or equivalently, in $\epsilon$).  
We consider a plasmas with $p_{th}=20$, and can approximate $\epsilon \approx p$.  
The fourth term in \cref{RadiationForce_x,RadiationForce_y} scales as second order in $p$, while the second and third terms are zeroth order in $p$.  
The first term is proportional to $dE/dt$, which is known to vanish in the relativistic regime compared with the fourth term.  
Keeping only the fourth term, we can express the radiation force as

\begin{multline}
\label{RadiationForceRElativistic}
    |\mathbf{F}_{R}|=\frac{2}{3}\alpha \epsilon p B_0^2
    \bigg[
    \frac{p^2}{\epsilon^2}+\frac{E_0^2}{B_0^2}-\frac{2}{B_0\epsilon}(E_xp_y-E_yp_x)\\-\frac{1}{\epsilon^2B_0^2}\big(p_xE_x+p_yE_y\big)^2
    \bigg]
\end{multline}
The first term is familiar from the synchrotron case, the second and 
fourth terms originate from pure electric-field radiation, and the 
third arises from the combined action of electric and magnetic fields. 
Note that both the first and second terms yield an azimuthally symmetric 
radiation force, whereas the third and fourth terms can break this 
symmetry depending on the values of $E_x$ and $E_y$.
In the limit of $E_0\ll B_0$, we can keep up to first order in $E_0/B_0$ and get the radiation force $|\mathbf{F}_{R}|$ divided with the pure magnetic field radiation force $|\mathbf{F}_{R,E=0}|$
\begin{equation}
\label{RadiationForce_EandB}
    \frac{|\mathbf{F}_{R}|}{|\mathbf{F}_{R,E=0|}}= 1-\frac{2\epsilon}{p^2 B_0}\Big(E_xp_y-E_yp_x\Big)
\end{equation}
This equation shows that the electric field can enhance and decrease the total radiation force depending on the values of the two components of the electric field.
To investigate how the electric field influences radiation, we begin our analysis of 
electric-field propagation in a magnetized plasma by considering the limit 
$E_0 \ll B_0$. The aim is to smoothly introduce the effect of radiation reaction due to 
the electric field on the azimuthally symmetric magnetic-field case studied in 
\cref{SynchrotronRadiation}, and to examine how the ring momentum anisotropy is modified. 
In \cref{Fig6}, we present simulation results of radiation reaction in the combined presence 
of electric and magnetic fields. In particular, we illustrate the behavior of the radiation 
force $|\mathbf{F}_{\mathrm{R}} f_0|$ within the initial plasma distribution. 
We adopt $p_{\mathrm{th}} = 20$ to ensure a relativistic temperature. 
The plasma density is chosen as $n_0 = 7 \times 10^{25}\,\mathrm{cm}^{-3}$, 
which corresponds to a rest-frame plasma frequency of $\omega_p = 0.0096$ 
(note that in the lab frame this frequency is reduced due to the 
$\omega_p/\gamma$ scaling). For the magnetic field, we set $B_0 = 0.05$, 
giving a cyclotron frequency of $\omega_c = 0.05$. Thus, we study the 
regime where $\omega_c \gg \omega_p$, implying that $\omega_c \sim \omega_{uh}$, 
with $\omega_{uh}$ denoting the upper hybrid frequency.

In \cref{Fig6}(a), we show the radiation reaction force in the case of a pure magnetic field, 
which serves primarily as a reference for comparison with the subsequent subfigures 
where the electric field modifies the force structure. In \cref{Fig6}(b), we include an 
electric field with $E_0 = 0.005$ and display the radiation reaction force at $t = 0$, when 
$E_x = E_y = E_0$. Comparing this result with \cref{Fig6}(a), two regions exhibit notable 
differences. The first is at large $p_x$ and negative $p_y$, where the second term in 
\cref{RadiationForce_EandB} becomes positive, leading to an enhanced radiation force. 
The second is at large positive $p_y$ and negative $p_x$, where the same term becomes 
negative, producing a weakened force, as seen in \cref{Fig6}(b).  
In the remaining regions of the $p_x$–$p_y$ plane, the behavior closely resembles 
the pure magnetic-field case. In particular, for combinations of large positive $p_x$ 
and $p_y$ or large negative $p_x$ and $p_y$, the contributions from the two parts of the 
second term in \cref{RadiationForce_EandB} effectively cancel.

In \cref{Fig6}(c), we consider $E_0 = 0.01$ and display the radiation reaction force at $t = 0$. 
The same qualitative effects as in \cref{Fig6}(b) are observed, but the modifications 
induced by the electric field are more pronounced since the field amplitude is twice 
as large, resulting in a doubled contribution from the second term in 
\cref{RadiationForce_EandB}.  
In \cref{Fig6}(d–e), we show the radiation force at $t = 0.6 t_R$ for $E_0 = 0.005$ 
and $E_0 = 0.01$, respectively. In contrast to \cref{Fig6}(b–c), the electric field 
components are no longer equal, but instead $E_x$ is negative while $E_y$ is positive, 
with differing magnitudes. This asymmetry shifts the regions in the $p_x$–$p_y$ plane 
where the second term of \cref{RadiationForce_EandB} contributes positively or negatively 
to the radiation force.

\begin{figure}
    \centering
    \includegraphics[width=8.5 cm, height=10 cm]{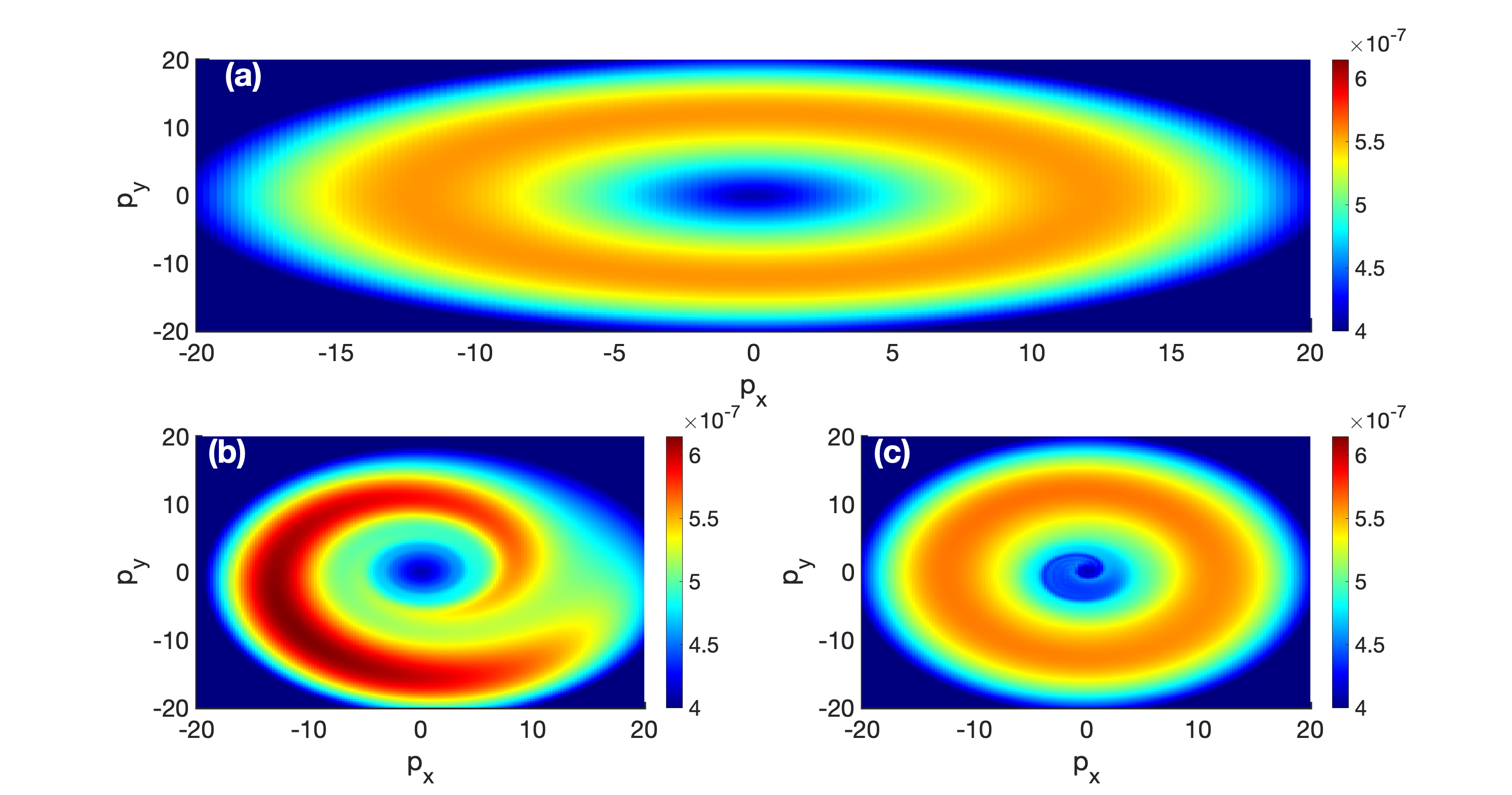}
    \caption{The plasma distribution at $t=0.6t_R$ is shown as follows: 
(a) pure magnetic field in plasma, 
(b) with an electric field $E_0=0.01$ and plasma density $n_0 = 7\times 10^{25}\,\text{cm}^{-3}$, 
and (c) with $E_0=0.01$ and a plasma density of $n_0 = 7\times 10^{27}\,\text{cm}^{-3}$.
 }
    \label{Fig7}
\end{figure}

To understand how the new momentum-shape of the total radiation force presented in \cref{Fig6} would affect the initial plasma distribution, we plot
 the plasma distribution after radiation at $t=0.6 t_R$ in \cref{Fig7}. In \cref{Fig7}(a), we plot the plasma distribution for the pure synchrotron case, which serves as a reference for the other subplots. In \cref{Fig7}(b), we display the plasma distribution in the presence of both electric and magnetic fields, with $B_0=0.05$, $E_0=0.01$, and plasma density $n_0 = 7\times 10^{25}\,\text{cm}^{-3}$, the same parameters used in \cref{Fig6}(c,e), i.e. we have $\omega_c\gg \omega_p$. As shown previously in \cref{Fig6}(c,e), two distinct regions emerge where the net radiation force is stronger or weaker compared to the pure synchrotron case. In the right half of the $p_x$–$p_y$ plane, the radiation force is enhanced due to the positive contribution of the electric field, causing some electrons to be pushed toward the weaker-radiation region in the left half of the plane. This redistribution can be confirmed by comparing the color map in \cref{Fig7}(b) with that in \cref{Fig7}(a): in the left half of the $p_x$–$p_y$ plane, the number of electrons increases due to the weaker radiation force, consistent with the behavior shown in \cref{Fig6}(c,e).

Increasing the plasma density raises the oscillation frequency of the electric field and we enter the regime where $\omega_p \gg \omega_c$. Consequently, during the radiation time $0.6 t_R$, the components of the electric field, $E_x$ and $E_y$, undergo many oscillation periods. This leads to the possibility that the second term of \cref{RadiationForce_EandB} contributes both positively and negatively to the net radiation force within the same regions of the $p_x$–$p_y$ plane. This effect is illustrated in \cref{Fig7}(c), where we use $E_0=0.01$, $B_0=0.05$, and $n_0 = 7\times 10^{27}\,\text{cm}^{-3}$. The resulting plasma distribution is very similar to the pure synchrotron case in \cref{Fig7}(a). The high frequency of the self-consistent electric field causes the positive and negative contributions of the electric field to the net radiation force to shift rapidly across the $p_x$–$p_y$ plane. As a result, different regions of momentum space experience alternating contributions at different times, which produces an almost symmetric plasma distribution. Moreover, the duration over which any given region of momentum space is influenced by these positive or negative contributions is very short, since the higher oscillation frequency causes the contributions to move quickly from one region to another.

Studying the time evolution of the ring radius in Fig.~\ref{Fig7}(c), we find that it closely resembles the case of a pure magnetic field. One might suspect that the moderate electric field amplitude of $E_0 = 0.01$, compared to $B_0 = 0.05$, is the reason why the time evolution of the ring radius appears almost unaffected by the presence of the electric field. However, this is not the case, as the rapid oscillations of the electric field suppress the influence of the azimuthally asymmetric term on the total radiation-reaction force. Increasing the electric field amplitude to $E_0 = 0.05$, while keeping the same magnetic field and plasma density, the time evolution of the ring radius remains essentially unchanged. This behavior can be explained using \cref{RadiationForceRElativistic}. By taking $E_0 = B_0$ and $p \sim \epsilon$, which is a valid approximation for a relativistic thermal plasma (as radiation primarily occurs at large momentum), we obtain

\begin{multline*}
    |\mathbf{F}_{R}|=\frac{2}{3}\alpha \epsilon p B_0^2
    \bigg[
    1+1-\frac{2}{B_0\epsilon}(E_xp_y-E_yp_x)\\-
    \frac{1}{\epsilon^2B_0^2}\big(p_xE_x+p_yE_y\big)^2
    \bigg]
\end{multline*}
By examining the time evolution of $E_x$ and $E_y$, we find that their behaviors closely coincide. Thus, we may approximate $E_y \approx E_x$. This leads to
\begin{equation*}
        \frac{1}{\epsilon^2B_0^2}\big(p_xE_x+p_yE_y\big)^2=    \frac{E_x^2}{B_0^2}
\end{equation*}
The maximum value of $E_x$ is $E_0$, which is equal to $B_0$. At the time when the electric field reaches its peak, we obtain

\begin{multline}
\label{Last_equation}
    |\mathbf{F}_{R}|=\frac{2}{3}\alpha \epsilon p B_0^2
    \bigg[
    1+1-\frac{2}{B_0\epsilon}(E_xp_y-E_yp_x)-1   \bigg]\\
 =\frac{2}{3}\alpha \epsilon p B_0^2
    \bigg[
    1-\frac{2}{B_0\epsilon}(E_xp_y-E_yp_x)  \bigg]\
\end{multline}
The second and fourth terms cancel each other when $E_x$ and $E_y$ attain their peak values. At the time when the electric field components have the value 0, the results in \cref{Last_equation} still hold as the second term that is $E_0^2/B_0^2= (E_x^2+E_y^2)/B_0^2$ would vanish together with the fourth term.  At intermediate times, the first term in \cref{Last_equation}  takes values between one and two. After applying these assumptions, only the third term of \cref{RadiationForceRElativistic} remains, whose effect is illustrated in Fig.~\ref{Fig6}(b--e) (note that we have used $p \sim \epsilon$). Thus, the conclusions derived for $E_{0} \ll B_{0}$ remain valid also in the case $E_{0} = B_{0}$.
Even for $E_0/B_0 =1$, the azimuthally asymmetric part of the radiation-reaction force does not alter the anisotropy formation, provided that the cyclotron frequency is much smaller than the upper hybrid frequency.
Due to the rapid oscillations of the electric field driven by the high upper hybrid frequency, the contribution of the azimuthally asymmetric term to the total radiation force shifts quickly in momentum space. This stands in contrast to the first term of \cref{Last_equation}, which remains constant over time. Consequently, the plasma distribution is only weakly affected, and the dynamics of the ring-momentum distribution remain largely unchanged.

\section{Discussion}
The main objective of this work is to investigate how the radiation reaction affect the momentum distribution of a plasma that is initial in equilibrium. We have used the Landau-Lifshitz force to model the recoil force of the radiation on the plasma. As the recoil force is non-conservative, the cooling of the plasma due to radiation becomes anisotropic in momentum-space, with stronger cooling on the region with stronger magnitude of the radiation reaction force. For the case of pure magnetic field it is shown that the azimuthal-free radiation reaction force lead to ring-momentum distribution. The time evolution characterizing the ring-momentum distribution is shown to agree well with the corresponding analytical solution.
Studying radiation due to strong magnetic field in anisotropic plasma, it is shown that introduction of initial anisotropy in momentum-space reduces the energy loss rate and as consequence the radius of the ring-momentum distribution become smaller. Due to the temperature imbalance between the \(x\)- and \(y\)-directions, the magnetic force induces a drift of particles in momentum space from the higher-temperature region to the lower-temperature one. This redistribution reduces the overall energy loss due to radiation.

Studying the radiation due to electric field, we found that the radiation reaction force is not azimuthal-free, i.e. the force is zero in certain regions in $p_x-p_y$-plane, while in other regions it is non-zero. In particular, the force is non-zero at the 90°–180° and 270°–360° sectors of the $p_x-p_y$-plane, while the remaining quadrants it is zero. The region where the force is non-zero experience a similar bunching of the electron due to the cooling effect of radiation. This leads to the creation of sector-dependent ring-momentum distribution with a similar time-dependent dynamics of the ring-radius as in the magnetic field-case. For the rest of the sectors in the $p_x-p_y$-plane, the plasma remains unaffected.

When both electric and magnetic fields are present in the plasma, the radiation reaction force can be decomposed into two contributions. The first is an azimuthally independent force that remains constant in time and drives the formation of a ring-momentum distribution. The second depends on the azimuthal angle, enhancing or reducing the total radiation force depending on the region of the $p_x$--$p_y$ plane. Since the second contribution is determined by the two components of the self-consistent electric field, which vary with time, the regions in momentum space where the total force is stronger or weaker continuously shift.  
The resulting anisotropy depends on whether the cyclotron frequency is near the upper hybrid frequency or not. When the cyclotron frequency is nearly resonant with the upper hybrid frequency, the electric-field components vary only slowly over the plasma cooling time. In this regime, the azimuthal-dependent force remains localized in the same region of momentum space long enough to significantly distort the plasma distribution, thereby breaking the azimuthal symmetry of the ring-momentum structure.  
In contrast, when the upper hybrid frequency is much larger than the cyclotron frequency, the rapid oscillations of the electric field components cause the positive and negative contributions of the azimuthal-dependent force to shift quickly across momentum space. As a result, the azimuthal symmetry of the ring-momentum anisotropy is preserved.

The physical parameters considered in this work, namely $B_0 = 0.05$, 
$E_0 = 0.01$, and the plasma density $n_0 = 7 \times 10^{25}$--$10^{27}\,
\mathrm{cm}^{-3}$, occur naturally in the magnetospheres of compact 
astrophysical objects \cite{Uzdensky2023}. In a laboratory context, it is 
anticipated that next-generation strong-field experiments will be capable 
of creating analogous astrophysical conditions \cite{Fiuza2023}.

The Vlasov system with a linearly polarized electric field and a constant 
magnetic field, including Ampère’s law and the radiation reaction force, 
has been solved using a two-step numerical scheme. The numerical results 
show good agreement with the analytical solution in the limit of a vanishing 
electric field. Our focus was to investigate how the propagation of the 
electric field in a magnetized plasma affects the development of the 
ring-momentum distribution. Thus, the physics related to the onset of 
instabilities (which require much longer timescales) arising after the 
development of anisotropies is beyond the scope of this work. For future 
studies, it would be of interest to investigate such instabilities, for 
example the electron cyclotron maser instability (see 
\cite{RingMomentum1,Instability,Instability1,Instability2,Instability3}). 
Furthermore, studying the development of anisotropies in momentum space 
driven by the electromagnetic field geometry would provide deeper insights 
into the nonlinear dynamics of magnetized plasmas and their associated 
radiation processes.

\label{DiscussionSection}

\section{ACKNOWLEDGMENT}
The author acknowledges support from the Knut and
Alice Wallenberg Foundation. A special thank to Gert
Brodin and David Reis for fruitful discussions.
 \bibliography{References}

\begin{thebibliography}{43}%
\makeatletter
\providecommand \@ifxundefined [1]{%
 \@ifx{#1\undefined}
}%
\providecommand \@ifnum [1]{%
 \ifnum #1\expandafter \@firstoftwo
 \else \expandafter \@secondoftwo
 \fi
}%
\providecommand \@ifx [1]{%
 \ifx #1\expandafter \@firstoftwo
 \else \expandafter \@secondoftwo
 \fi
}%
\providecommand \natexlab [1]{#1}%
\providecommand \enquote  [1]{``#1''}%
\providecommand \bibnamefont  [1]{#1}%
\providecommand \bibfnamefont [1]{#1}%
\providecommand \citenamefont [1]{#1}%
\providecommand \href@noop [0]{\@secondoftwo}%
\providecommand \href [0]{\begingroup \@sanitize@url \@href}%
\providecommand \@href[1]{\@@startlink{#1}\@@href}%
\providecommand \@@href[1]{\endgroup#1\@@endlink}%
\providecommand \@sanitize@url [0]{\catcode `\\12\catcode `\$12\catcode `\&12\catcode `\#12\catcode `\^12\catcode `\_12\catcode `\%12\relax}%
\providecommand \@@startlink[1]{}%
\providecommand \@@endlink[0]{}%
\providecommand \url  [0]{\begingroup\@sanitize@url \@url }%
\providecommand \@url [1]{\endgroup\@href {#1}{\urlprefix }}%
\providecommand \urlprefix  [0]{URL }%
\providecommand \Eprint [0]{\href }%
\providecommand \doibase [0]{https://doi.org/}%
\providecommand \selectlanguage [0]{\@gobble}%
\providecommand \bibinfo  [0]{\@secondoftwo}%
\providecommand \bibfield  [0]{\@secondoftwo}%
\providecommand \translation [1]{[#1]}%
\providecommand \BibitemOpen [0]{}%
\providecommand \bibitemStop [0]{}%
\providecommand \bibitemNoStop [0]{.\EOS\space}%
\providecommand \EOS [0]{\spacefactor3000\relax}%
\providecommand \BibitemShut  [1]{\csname bibitem#1\endcsname}%
\let\auto@bib@innerbib\@empty
\bibitem [{\citenamefont {Qu}\ \emph {et~al.}(2021)\citenamefont {Qu}, \citenamefont {Meuren},\ and\ \citenamefont {Fisch}}]{QEDPlasma1}%
  \BibitemOpen
  \bibfield  {author} {\bibinfo {author} {\bibfnamefont {K.}~\bibnamefont {Qu}}, \bibinfo {author} {\bibfnamefont {S.}~\bibnamefont {Meuren}},\ and\ \bibinfo {author} {\bibfnamefont {N.~J.}\ \bibnamefont {Fisch}},\ }\bibfield  {title} {\bibinfo {title} {Signature of collective plasma effects in beam-driven qed cascades},\ }\href@noop {} {\bibfield  {journal} {\bibinfo  {journal} {Physical review letters}\ }\textbf {\bibinfo {volume} {127}},\ \bibinfo {pages} {095001} (\bibinfo {year} {2021})}\BibitemShut {NoStop}%
\bibitem [{\citenamefont {Griffith}\ \emph {et~al.}(2024)\citenamefont {Griffith}, \citenamefont {Qu},\ and\ \citenamefont {Fisch}}]{QEDPlasma2}%
  \BibitemOpen
  \bibfield  {author} {\bibinfo {author} {\bibfnamefont {A.}~\bibnamefont {Griffith}}, \bibinfo {author} {\bibfnamefont {K.}~\bibnamefont {Qu}},\ and\ \bibinfo {author} {\bibfnamefont {N.~J.}\ \bibnamefont {Fisch}},\ }\bibfield  {title} {\bibinfo {title} {Radiation reaction kinetics and collective qed signatures},\ }\href@noop {} {\bibfield  {journal} {\bibinfo  {journal} {Physics of Plasmas}\ }\textbf {\bibinfo {volume} {31}} (\bibinfo {year} {2024})}\BibitemShut {NoStop}%
\bibitem [{\citenamefont {Qu}\ \emph {et~al.}(2022)\citenamefont {Qu}, \citenamefont {Meuren},\ and\ \citenamefont {Fisch}}]{QEDPlasma3}%
  \BibitemOpen
  \bibfield  {author} {\bibinfo {author} {\bibfnamefont {K.}~\bibnamefont {Qu}}, \bibinfo {author} {\bibfnamefont {S.}~\bibnamefont {Meuren}},\ and\ \bibinfo {author} {\bibfnamefont {N.~J.}\ \bibnamefont {Fisch}},\ }\bibfield  {title} {\bibinfo {title} {Collective plasma effects of electron--positron pairs in beam-driven qed cascades},\ }\href@noop {} {\bibfield  {journal} {\bibinfo  {journal} {Physics of Plasmas}\ }\textbf {\bibinfo {volume} {29}} (\bibinfo {year} {2022})}\BibitemShut {NoStop}%
\bibitem [{\citenamefont {Bilbao}\ and\ \citenamefont {Silva}(2023)}]{RingMomentum1}%
  \BibitemOpen
  \bibfield  {author} {\bibinfo {author} {\bibfnamefont {P.~J.}\ \bibnamefont {Bilbao}}\ and\ \bibinfo {author} {\bibfnamefont {L.~O.}\ \bibnamefont {Silva}},\ }\bibfield  {title} {\bibinfo {title} {Radiation reaction cooling as a source of anisotropic momentum distributions with inverted populations},\ }\href@noop {} {\bibfield  {journal} {\bibinfo  {journal} {Physical Review Letters}\ }\textbf {\bibinfo {volume} {130}},\ \bibinfo {pages} {165101} (\bibinfo {year} {2023})}\BibitemShut {NoStop}%
\bibitem [{\citenamefont {Bilbao}\ \emph {et~al.}(2024)\citenamefont {Bilbao}, \citenamefont {Ewart}, \citenamefont {Assun{\c{c}}ao}, \citenamefont {Silva},\ and\ \citenamefont {Silva}}]{RingMomentum2}%
  \BibitemOpen
  \bibfield  {author} {\bibinfo {author} {\bibfnamefont {P.~J.}\ \bibnamefont {Bilbao}}, \bibinfo {author} {\bibfnamefont {R.~J.}\ \bibnamefont {Ewart}}, \bibinfo {author} {\bibfnamefont {F.}~\bibnamefont {Assun{\c{c}}ao}}, \bibinfo {author} {\bibfnamefont {T.}~\bibnamefont {Silva}},\ and\ \bibinfo {author} {\bibfnamefont {L.~O.}\ \bibnamefont {Silva}},\ }\bibfield  {title} {\bibinfo {title} {Ring momentum distributions as a general feature of vlasov dynamics in the synchrotron dominated regime},\ }\href@noop {} {\bibfield  {journal} {\bibinfo  {journal} {Physics of Plasmas}\ }\textbf {\bibinfo {volume} {31}} (\bibinfo {year} {2024})}\BibitemShut {NoStop}%
\bibitem [{\citenamefont {Bulanov}\ \emph {et~al.}(2024)\citenamefont {Bulanov}, \citenamefont {Grittani}, \citenamefont {Shaisultanov}, \citenamefont {Esirkepov}, \citenamefont {Ridgers}, \citenamefont {Bulanov}, \citenamefont {Russell},\ and\ \citenamefont {Thomas}}]{bulanov2024energy}%
  \BibitemOpen
  \bibfield  {author} {\bibinfo {author} {\bibfnamefont {S.}~\bibnamefont {Bulanov}}, \bibinfo {author} {\bibfnamefont {G.}~\bibnamefont {Grittani}}, \bibinfo {author} {\bibfnamefont {R.}~\bibnamefont {Shaisultanov}}, \bibinfo {author} {\bibfnamefont {T.~Z.}\ \bibnamefont {Esirkepov}}, \bibinfo {author} {\bibfnamefont {C.}~\bibnamefont {Ridgers}}, \bibinfo {author} {\bibfnamefont {S.}~\bibnamefont {Bulanov}}, \bibinfo {author} {\bibfnamefont {B.}~\bibnamefont {Russell}},\ and\ \bibinfo {author} {\bibfnamefont {A.}~\bibnamefont {Thomas}},\ }\bibfield  {title} {\bibinfo {title} {On the energy spectrum evolution of electrons undergoing radiation cooling},\ }\href@noop {} {\bibfield  {journal} {\bibinfo  {journal} {Fundamental Plasma Physics}\ }\textbf {\bibinfo {volume} {9}},\ \bibinfo {pages} {100036} (\bibinfo {year} {2024})}\BibitemShut {NoStop}%
\bibitem [{\citenamefont {Chen}\ and\ \citenamefont {Fiuza}(2023)}]{Fiuza2023}%
  \BibitemOpen
  \bibfield  {author} {\bibinfo {author} {\bibfnamefont {H.}~\bibnamefont {Chen}}\ and\ \bibinfo {author} {\bibfnamefont {F.}~\bibnamefont {Fiuza}},\ }\bibfield  {title} {\bibinfo {title} {Perspectives on relativistic electron--positron pair plasma experiments of astrophysical relevance using high-power lasers},\ }\href@noop {} {\bibfield  {journal} {\bibinfo  {journal} {Physics of Plasmas}\ }\textbf {\bibinfo {volume} {30}} (\bibinfo {year} {2023})}\BibitemShut {NoStop}%
\bibitem [{\citenamefont {Gong}\ \emph {et~al.}(2019)\citenamefont {Gong}, \citenamefont {Mackenroth}, \citenamefont {Yan},\ and\ \citenamefont {Arefiev}}]{gong2019radiation}%
  \BibitemOpen
  \bibfield  {author} {\bibinfo {author} {\bibfnamefont {Z.}~\bibnamefont {Gong}}, \bibinfo {author} {\bibfnamefont {F.}~\bibnamefont {Mackenroth}}, \bibinfo {author} {\bibfnamefont {X.}~\bibnamefont {Yan}},\ and\ \bibinfo {author} {\bibfnamefont {A.}~\bibnamefont {Arefiev}},\ }\bibfield  {title} {\bibinfo {title} {Radiation reaction as an energy enhancement mechanism for laser-irradiated electrons in a strong plasma magnetic field},\ }\href@noop {} {\bibfield  {journal} {\bibinfo  {journal} {Scientific reports}\ }\textbf {\bibinfo {volume} {9}},\ \bibinfo {pages} {17181} (\bibinfo {year} {2019})}\BibitemShut {NoStop}%
\bibitem [{\citenamefont {Comisso}\ and\ \citenamefont {Sironi}(2021)}]{Lorenzo2021}%
  \BibitemOpen
  \bibfield  {author} {\bibinfo {author} {\bibfnamefont {L.}~\bibnamefont {Comisso}}\ and\ \bibinfo {author} {\bibfnamefont {L.}~\bibnamefont {Sironi}},\ }\bibfield  {title} {\bibinfo {title} {Pitch-angle anisotropy controls particle acceleration and cooling in radiative relativistic plasma turbulence},\ }\href@noop {} {\bibfield  {journal} {\bibinfo  {journal} {Physical Review Letters}\ }\textbf {\bibinfo {volume} {127}},\ \bibinfo {pages} {255102} (\bibinfo {year} {2021})}\BibitemShut {NoStop}%
\bibitem [{\citenamefont {Zhdankin}\ \emph {et~al.}(2023)\citenamefont {Zhdankin}, \citenamefont {Kunz},\ and\ \citenamefont {Uzdensky}}]{Uzdensky2023}%
  \BibitemOpen
  \bibfield  {author} {\bibinfo {author} {\bibfnamefont {V.}~\bibnamefont {Zhdankin}}, \bibinfo {author} {\bibfnamefont {M.~W.}\ \bibnamefont {Kunz}},\ and\ \bibinfo {author} {\bibfnamefont {D.~A.}\ \bibnamefont {Uzdensky}},\ }\bibfield  {title} {\bibinfo {title} {Synchrotron firehose instability},\ }\href@noop {} {\bibfield  {journal} {\bibinfo  {journal} {The Astrophysical Journal}\ }\textbf {\bibinfo {volume} {944}},\ \bibinfo {pages} {24} (\bibinfo {year} {2023})}\BibitemShut {NoStop}%
\bibitem [{\citenamefont {Liseykina}\ \emph {et~al.}(2016)\citenamefont {Liseykina}, \citenamefont {Popruzhenko},\ and\ \citenamefont {Macchi}}]{liseykina2016inverse}%
  \BibitemOpen
  \bibfield  {author} {\bibinfo {author} {\bibfnamefont {T.}~\bibnamefont {Liseykina}}, \bibinfo {author} {\bibfnamefont {S.}~\bibnamefont {Popruzhenko}},\ and\ \bibinfo {author} {\bibfnamefont {A.}~\bibnamefont {Macchi}},\ }\bibfield  {title} {\bibinfo {title} {Inverse faraday effect driven by radiation friction},\ }\href@noop {} {\bibfield  {journal} {\bibinfo  {journal} {New Journal of Physics}\ }\textbf {\bibinfo {volume} {18}},\ \bibinfo {pages} {072001} (\bibinfo {year} {2016})}\BibitemShut {NoStop}%
\bibitem [{\citenamefont {Zhang}\ \emph {et~al.}(2020)\citenamefont {Zhang}, \citenamefont {Bulanov}, \citenamefont {Seipt}, \citenamefont {Arefiev},\ and\ \citenamefont {Thomas}}]{zhang2020relativistic}%
  \BibitemOpen
  \bibfield  {author} {\bibinfo {author} {\bibfnamefont {P.}~\bibnamefont {Zhang}}, \bibinfo {author} {\bibfnamefont {S.}~\bibnamefont {Bulanov}}, \bibinfo {author} {\bibfnamefont {D.}~\bibnamefont {Seipt}}, \bibinfo {author} {\bibfnamefont {A.}~\bibnamefont {Arefiev}},\ and\ \bibinfo {author} {\bibfnamefont {A.}~\bibnamefont {Thomas}},\ }\bibfield  {title} {\bibinfo {title} {Relativistic plasma physics in supercritical fields},\ }\href@noop {} {\bibfield  {journal} {\bibinfo  {journal} {Physics of Plasmas}\ }\textbf {\bibinfo {volume} {27}} (\bibinfo {year} {2020})}\BibitemShut {NoStop}%
\bibitem [{\citenamefont {Chen}\ \emph {et~al.}(2022)\citenamefont {Chen}, \citenamefont {Meuren}, \citenamefont {Gerstmayr}, \citenamefont {Yakimenko}, \citenamefont {Bucksbaum},\ and\ \citenamefont {Reis}}]{E320}%
  \BibitemOpen
  \bibfield  {author} {\bibinfo {author} {\bibfnamefont {Z.}~\bibnamefont {Chen}}, \bibinfo {author} {\bibfnamefont {S.}~\bibnamefont {Meuren}}, \bibinfo {author} {\bibfnamefont {E.}~\bibnamefont {Gerstmayr}}, \bibinfo {author} {\bibfnamefont {V.}~\bibnamefont {Yakimenko}}, \bibinfo {author} {\bibfnamefont {P.~H.}\ \bibnamefont {Bucksbaum}},\ and\ \bibinfo {author} {\bibfnamefont {D.~A.}\ \bibnamefont {Reis}},\ }\bibfield  {title} {\bibinfo {title} {Preparation of strong-field qed experiments at facet-ii},\ }in\ \href@noop {} {\emph {\bibinfo {booktitle} {High Intensity Lasers and High Field Phenomena}}}\ (\bibinfo {organization} {Optica Publishing Group},\ \bibinfo {year} {2022})\ pp.\ \bibinfo {pages} {HF4B--6}\BibitemShut {NoStop}%
\bibitem [{\citenamefont {Burke}\ \emph {et~al.}(1997)\citenamefont {Burke}, \citenamefont {Field}, \citenamefont {Horton-Smith}, \citenamefont {Spencer}, \citenamefont {Walz}, \citenamefont {Berridge}, \citenamefont {Bugg}, \citenamefont {Shmakov}, \citenamefont {Weidemann}, \citenamefont {Bula}, \citenamefont {McDonald}, \citenamefont {Prebys}, \citenamefont {Bamber}, \citenamefont {Boege}, \citenamefont {Koffas}, \citenamefont {Kotseroglou}, \citenamefont {Melissinos}, \citenamefont {Meyerhofer}, \citenamefont {Reis},\ and\ \citenamefont {Ragg}}]{SLAC1997}%
  \BibitemOpen
  \bibfield  {author} {\bibinfo {author} {\bibfnamefont {D.~L.}\ \bibnamefont {Burke}}, \bibinfo {author} {\bibfnamefont {R.~C.}\ \bibnamefont {Field}}, \bibinfo {author} {\bibfnamefont {G.}~\bibnamefont {Horton-Smith}}, \bibinfo {author} {\bibfnamefont {J.~E.}\ \bibnamefont {Spencer}}, \bibinfo {author} {\bibfnamefont {D.}~\bibnamefont {Walz}}, \bibinfo {author} {\bibfnamefont {S.~C.}\ \bibnamefont {Berridge}}, \bibinfo {author} {\bibfnamefont {W.~M.}\ \bibnamefont {Bugg}}, \bibinfo {author} {\bibfnamefont {K.}~\bibnamefont {Shmakov}}, \bibinfo {author} {\bibfnamefont {A.~W.}\ \bibnamefont {Weidemann}}, \bibinfo {author} {\bibfnamefont {C.}~\bibnamefont {Bula}}, \bibinfo {author} {\bibfnamefont {K.~T.}\ \bibnamefont {McDonald}}, \bibinfo {author} {\bibfnamefont {E.~J.}\ \bibnamefont {Prebys}}, \bibinfo {author} {\bibfnamefont {C.}~\bibnamefont {Bamber}}, \bibinfo {author} {\bibfnamefont {S.~J.}\ \bibnamefont {Boege}}, \bibinfo {author} {\bibfnamefont {T.}~\bibnamefont {Koffas}}, \bibinfo {author}
  {\bibfnamefont {T.}~\bibnamefont {Kotseroglou}}, \bibinfo {author} {\bibfnamefont {A.~C.}\ \bibnamefont {Melissinos}}, \bibinfo {author} {\bibfnamefont {D.~D.}\ \bibnamefont {Meyerhofer}}, \bibinfo {author} {\bibfnamefont {D.~A.}\ \bibnamefont {Reis}},\ and\ \bibinfo {author} {\bibfnamefont {W.}~\bibnamefont {Ragg}},\ }\bibfield  {title} {\bibinfo {title} {Positron production in multiphoton light-by-light scattering},\ }\href {https://doi.org/10.1103/PhysRevLett.79.1626} {\bibfield  {journal} {\bibinfo  {journal} {Phys. Rev. Lett.}\ }\textbf {\bibinfo {volume} {79}},\ \bibinfo {pages} {1626} (\bibinfo {year} {1997})}\BibitemShut {NoStop}%
\bibitem [{\citenamefont {Cole}\ \emph {et~al.}(2018)\citenamefont {Cole}, \citenamefont {Behm}, \citenamefont {Gerstmayr}, \citenamefont {Blackburn}, \citenamefont {Wood}, \citenamefont {Baird}, \citenamefont {Duff}, \citenamefont {Harvey}, \citenamefont {Ilderton}, \citenamefont {Joglekar}, \citenamefont {Krushelnick}, \citenamefont {Kuschel}, \citenamefont {Marklund}, \citenamefont {McKenna}, \citenamefont {Murphy}, \citenamefont {Poder}, \citenamefont {Ridgers}, \citenamefont {Samarin}, \citenamefont {Sarri}, \citenamefont {Symes}, \citenamefont {Thomas}, \citenamefont {Warwick}, \citenamefont {Zepf}, \citenamefont {Najmudin},\ and\ \citenamefont {Mangles}}]{Rutherford}%
  \BibitemOpen
  \bibfield  {author} {\bibinfo {author} {\bibfnamefont {J.~M.}\ \bibnamefont {Cole}}, \bibinfo {author} {\bibfnamefont {K.~T.}\ \bibnamefont {Behm}}, \bibinfo {author} {\bibfnamefont {E.}~\bibnamefont {Gerstmayr}}, \bibinfo {author} {\bibfnamefont {T.~G.}\ \bibnamefont {Blackburn}}, \bibinfo {author} {\bibfnamefont {J.~C.}\ \bibnamefont {Wood}}, \bibinfo {author} {\bibfnamefont {C.~D.}\ \bibnamefont {Baird}}, \bibinfo {author} {\bibfnamefont {M.~J.}\ \bibnamefont {Duff}}, \bibinfo {author} {\bibfnamefont {C.}~\bibnamefont {Harvey}}, \bibinfo {author} {\bibfnamefont {A.}~\bibnamefont {Ilderton}}, \bibinfo {author} {\bibfnamefont {A.~S.}\ \bibnamefont {Joglekar}}, \bibinfo {author} {\bibfnamefont {K.}~\bibnamefont {Krushelnick}}, \bibinfo {author} {\bibfnamefont {S.}~\bibnamefont {Kuschel}}, \bibinfo {author} {\bibfnamefont {M.}~\bibnamefont {Marklund}}, \bibinfo {author} {\bibfnamefont {P.}~\bibnamefont {McKenna}}, \bibinfo {author} {\bibfnamefont {C.~D.}\ \bibnamefont {Murphy}}, \bibinfo {author}
  {\bibfnamefont {K.}~\bibnamefont {Poder}}, \bibinfo {author} {\bibfnamefont {C.~P.}\ \bibnamefont {Ridgers}}, \bibinfo {author} {\bibfnamefont {G.~M.}\ \bibnamefont {Samarin}}, \bibinfo {author} {\bibfnamefont {G.}~\bibnamefont {Sarri}}, \bibinfo {author} {\bibfnamefont {D.~R.}\ \bibnamefont {Symes}}, \bibinfo {author} {\bibfnamefont {A.~G.~R.}\ \bibnamefont {Thomas}}, \bibinfo {author} {\bibfnamefont {J.}~\bibnamefont {Warwick}}, \bibinfo {author} {\bibfnamefont {M.}~\bibnamefont {Zepf}}, \bibinfo {author} {\bibfnamefont {Z.}~\bibnamefont {Najmudin}},\ and\ \bibinfo {author} {\bibfnamefont {S.~P.~D.}\ \bibnamefont {Mangles}},\ }\bibfield  {title} {\bibinfo {title} {Experimental evidence of radiation reaction in the collision of a high-intensity laser pulse with a laser-wakefield accelerated electron beam},\ }\href {https://doi.org/10.1103/PhysRevX.8.011020} {\bibfield  {journal} {\bibinfo  {journal} {Phys. Rev. X}\ }\textbf {\bibinfo {volume} {8}},\ \bibinfo {pages} {011020} (\bibinfo {year}
  {2018})}\BibitemShut {NoStop}%
\bibitem [{\citenamefont {Fedotov}\ \emph {et~al.}(2023)\citenamefont {Fedotov}, \citenamefont {Ilderton}, \citenamefont {Karbstein}, \citenamefont {King}, \citenamefont {Seipt}, \citenamefont {Taya},\ and\ \citenamefont {Torgrimsson}}]{QED-review1}%
  \BibitemOpen
  \bibfield  {author} {\bibinfo {author} {\bibfnamefont {A.}~\bibnamefont {Fedotov}}, \bibinfo {author} {\bibfnamefont {A.}~\bibnamefont {Ilderton}}, \bibinfo {author} {\bibfnamefont {F.}~\bibnamefont {Karbstein}}, \bibinfo {author} {\bibfnamefont {B.}~\bibnamefont {King}}, \bibinfo {author} {\bibfnamefont {D.}~\bibnamefont {Seipt}}, \bibinfo {author} {\bibfnamefont {H.}~\bibnamefont {Taya}},\ and\ \bibinfo {author} {\bibfnamefont {G.}~\bibnamefont {Torgrimsson}},\ }\bibfield  {title} {\bibinfo {title} {Advances in qed with intense background fields},\ }\href@noop {} {\bibfield  {journal} {\bibinfo  {journal} {Physics Reports}\ }\textbf {\bibinfo {volume} {1010}},\ \bibinfo {pages} {1} (\bibinfo {year} {2023})}\BibitemShut {NoStop}%
\bibitem [{\citenamefont {Abramowicz}\ \emph {et~al.}(2021)\citenamefont {Abramowicz}, \citenamefont {Acosta}, \citenamefont {Altarelli}, \citenamefont {Assmann}, \citenamefont {Bai}, \citenamefont {Behnke}, \citenamefont {Benhammou}, \citenamefont {Blackburn}, \citenamefont {Boogert}, \citenamefont {Borysov} \emph {et~al.}}]{QED-review2}%
  \BibitemOpen
  \bibfield  {author} {\bibinfo {author} {\bibfnamefont {H.}~\bibnamefont {Abramowicz}}, \bibinfo {author} {\bibfnamefont {U.}~\bibnamefont {Acosta}}, \bibinfo {author} {\bibfnamefont {M.}~\bibnamefont {Altarelli}}, \bibinfo {author} {\bibfnamefont {R.}~\bibnamefont {Assmann}}, \bibinfo {author} {\bibfnamefont {Z.}~\bibnamefont {Bai}}, \bibinfo {author} {\bibfnamefont {T.}~\bibnamefont {Behnke}}, \bibinfo {author} {\bibfnamefont {Y.}~\bibnamefont {Benhammou}}, \bibinfo {author} {\bibfnamefont {T.}~\bibnamefont {Blackburn}}, \bibinfo {author} {\bibfnamefont {S.}~\bibnamefont {Boogert}}, \bibinfo {author} {\bibfnamefont {O.}~\bibnamefont {Borysov}}, \emph {et~al.},\ }\bibfield  {title} {\bibinfo {title} {Conceptual design report for the luxe experiment},\ }\href@noop {} {\bibfield  {journal} {\bibinfo  {journal} {The European Physical Journal Special Topics}\ }\textbf {\bibinfo {volume} {230}},\ \bibinfo {pages} {2445} (\bibinfo {year} {2021})}\BibitemShut {NoStop}%
\bibitem [{\citenamefont {Gonoskov}\ \emph {et~al.}(2022)\citenamefont {Gonoskov}, \citenamefont {Blackburn}, \citenamefont {Marklund},\ and\ \citenamefont {Bulanov}}]{QED-review3}%
  \BibitemOpen
  \bibfield  {author} {\bibinfo {author} {\bibfnamefont {A.}~\bibnamefont {Gonoskov}}, \bibinfo {author} {\bibfnamefont {T.~G.}\ \bibnamefont {Blackburn}}, \bibinfo {author} {\bibfnamefont {M.}~\bibnamefont {Marklund}},\ and\ \bibinfo {author} {\bibfnamefont {S.~S.}\ \bibnamefont {Bulanov}},\ }\bibfield  {title} {\bibinfo {title} {Charged particle motion and radiation in strong electromagnetic fields},\ }\href {https://doi.org/10.1103/RevModPhys.94.045001} {\bibfield  {journal} {\bibinfo  {journal} {Rev. Mod. Phys.}\ }\textbf {\bibinfo {volume} {94}},\ \bibinfo {pages} {045001} (\bibinfo {year} {2022})}\BibitemShut {NoStop}%
\bibitem [{\citenamefont {Ford}\ and\ \citenamefont {O'connell}(1991)}]{Runaway}%
  \BibitemOpen
  \bibfield  {author} {\bibinfo {author} {\bibfnamefont {G.}~\bibnamefont {Ford}}\ and\ \bibinfo {author} {\bibfnamefont {R.}~\bibnamefont {O'connell}},\ }\bibfield  {title} {\bibinfo {title} {Radiation reaction in electrodynamics and the elimination of runaway solutions},\ }\href@noop {} {\bibfield  {journal} {\bibinfo  {journal} {Physics Letters A}\ }\textbf {\bibinfo {volume} {157}},\ \bibinfo {pages} {217} (\bibinfo {year} {1991})}\BibitemShut {NoStop}%
\bibitem [{\citenamefont {Landau}(2013)}]{Landau_Lif}%
  \BibitemOpen
  \bibfield  {author} {\bibinfo {author} {\bibfnamefont {L.~D.}\ \bibnamefont {Landau}},\ }\href@noop {} {\emph {\bibinfo {title} {The classical theory of fields}}},\ Vol.~\bibinfo {volume} {2}\ (\bibinfo  {publisher} {Elsevier},\ \bibinfo {year} {2013})\BibitemShut {NoStop}%
\bibitem [{\citenamefont {Bulanov}\ \emph {et~al.}(2011)\citenamefont {Bulanov}, \citenamefont {Esirkepov}, \citenamefont {Kando}, \citenamefont {Koga},\ and\ \citenamefont {Bulanov}}]{Bulanov}%
  \BibitemOpen
  \bibfield  {author} {\bibinfo {author} {\bibfnamefont {S.~V.}\ \bibnamefont {Bulanov}}, \bibinfo {author} {\bibfnamefont {T.~Z.}\ \bibnamefont {Esirkepov}}, \bibinfo {author} {\bibfnamefont {M.}~\bibnamefont {Kando}}, \bibinfo {author} {\bibfnamefont {J.~K.}\ \bibnamefont {Koga}},\ and\ \bibinfo {author} {\bibfnamefont {S.~S.}\ \bibnamefont {Bulanov}},\ }\bibfield  {title} {\bibinfo {title} {Lorentz-abraham-dirac versus landau-lifshitz radiation friction force in the ultrarelativistic electron interaction with electromagnetic wave (exact solutions)},\ }\href {https://doi.org/10.1103/PhysRevE.84.056605} {\bibfield  {journal} {\bibinfo  {journal} {Phys. Rev. E}\ }\textbf {\bibinfo {volume} {84}},\ \bibinfo {pages} {056605} (\bibinfo {year} {2011})}\BibitemShut {NoStop}%
\bibitem [{\citenamefont {Piazza}(2008)}]{piazza2008exact}%
  \BibitemOpen
  \bibfield  {author} {\bibinfo {author} {\bibfnamefont {A.~D.}\ \bibnamefont {Piazza}},\ }\bibfield  {title} {\bibinfo {title} {Exact solution of the landau-lifshitz equation in a plane wave},\ }\href@noop {} {\bibfield  {journal} {\bibinfo  {journal} {Letters in Mathematical Physics}\ }\textbf {\bibinfo {volume} {83}},\ \bibinfo {pages} {305} (\bibinfo {year} {2008})}\BibitemShut {NoStop}%
\bibitem [{\citenamefont {Vranic}\ \emph {et~al.}(2016{\natexlab{a}})\citenamefont {Vranic}, \citenamefont {Martins}, \citenamefont {Fonseca},\ and\ \citenamefont {Silva}}]{Silva}%
  \BibitemOpen
  \bibfield  {author} {\bibinfo {author} {\bibfnamefont {M.}~\bibnamefont {Vranic}}, \bibinfo {author} {\bibfnamefont {J.~L.}\ \bibnamefont {Martins}}, \bibinfo {author} {\bibfnamefont {R.~A.}\ \bibnamefont {Fonseca}},\ and\ \bibinfo {author} {\bibfnamefont {L.~O.}\ \bibnamefont {Silva}},\ }\bibfield  {title} {\bibinfo {title} {Classical radiation reaction in particle-in-cell simulations},\ }\href@noop {} {\bibfield  {journal} {\bibinfo  {journal} {Computer Physics Communications}\ }\textbf {\bibinfo {volume} {204}},\ \bibinfo {pages} {141} (\bibinfo {year} {2016}{\natexlab{a}})}\BibitemShut {NoStop}%
\bibitem [{\citenamefont {Wallin}\ \emph {et~al.}(2017)\citenamefont {Wallin}, \citenamefont {Gonoskov}, \citenamefont {Harvey}, \citenamefont {Lundh},\ and\ \citenamefont {Marklund}}]{Wallin}%
  \BibitemOpen
  \bibfield  {author} {\bibinfo {author} {\bibfnamefont {E.}~\bibnamefont {Wallin}}, \bibinfo {author} {\bibfnamefont {A.}~\bibnamefont {Gonoskov}}, \bibinfo {author} {\bibfnamefont {C.}~\bibnamefont {Harvey}}, \bibinfo {author} {\bibfnamefont {O.}~\bibnamefont {Lundh}},\ and\ \bibinfo {author} {\bibfnamefont {M.}~\bibnamefont {Marklund}},\ }\bibfield  {title} {\bibinfo {title} {Ultra-intense laser pulses in near-critical underdense plasmas--radiation reaction and energy partitioning},\ }\href@noop {} {\bibfield  {journal} {\bibinfo  {journal} {Journal of Plasma Physics}\ }\textbf {\bibinfo {volume} {83}} (\bibinfo {year} {2017})}\BibitemShut {NoStop}%
\bibitem [{\citenamefont {Vranic}\ \emph {et~al.}(2016{\natexlab{b}})\citenamefont {Vranic}, \citenamefont {Grismayer}, \citenamefont {Fonseca},\ and\ \citenamefont {Silva}}]{Silva_Q}%
  \BibitemOpen
  \bibfield  {author} {\bibinfo {author} {\bibfnamefont {M.}~\bibnamefont {Vranic}}, \bibinfo {author} {\bibfnamefont {T.}~\bibnamefont {Grismayer}}, \bibinfo {author} {\bibfnamefont {R.~A.}\ \bibnamefont {Fonseca}},\ and\ \bibinfo {author} {\bibfnamefont {L.~O.}\ \bibnamefont {Silva}},\ }\bibfield  {title} {\bibinfo {title} {Quantum radiation reaction in head-on laser-electron beam interaction},\ }\href@noop {} {\bibfield  {journal} {\bibinfo  {journal} {New Journal of Physics}\ }\textbf {\bibinfo {volume} {18}},\ \bibinfo {pages} {073035} (\bibinfo {year} {2016}{\natexlab{b}})}\BibitemShut {NoStop}%
\bibitem [{\citenamefont {Hakim}\ and\ \citenamefont {Mangeney}(1968)}]{Hakim}%
  \BibitemOpen
  \bibfield  {author} {\bibinfo {author} {\bibfnamefont {R.}~\bibnamefont {Hakim}}\ and\ \bibinfo {author} {\bibfnamefont {A.}~\bibnamefont {Mangeney}},\ }\bibfield  {title} {\bibinfo {title} {Relativistic kinetic equations including radiation effects. i. vlasov approximation},\ }\href@noop {} {\bibfield  {journal} {\bibinfo  {journal} {Journal of Mathematical Physics}\ }\textbf {\bibinfo {volume} {9}},\ \bibinfo {pages} {116} (\bibinfo {year} {1968})}\BibitemShut {NoStop}%
\bibitem [{\citenamefont {Burton}\ and\ \citenamefont {Noble}(2014)}]{Burton}%
  \BibitemOpen
  \bibfield  {author} {\bibinfo {author} {\bibfnamefont {D.~A.}\ \bibnamefont {Burton}}\ and\ \bibinfo {author} {\bibfnamefont {A.}~\bibnamefont {Noble}},\ }\bibfield  {title} {\bibinfo {title} {Aspects of electromagnetic radiation reaction in strong fields},\ }\href@noop {} {\bibfield  {journal} {\bibinfo  {journal} {Contemporary Physics}\ }\textbf {\bibinfo {volume} {55}},\ \bibinfo {pages} {110} (\bibinfo {year} {2014})}\BibitemShut {NoStop}%
\bibitem [{\citenamefont {Kunze}\ and\ \citenamefont {Rendall}(2001)}]{Kunze}%
  \BibitemOpen
  \bibfield  {author} {\bibinfo {author} {\bibfnamefont {M.}~\bibnamefont {Kunze}}\ and\ \bibinfo {author} {\bibfnamefont {A.~D.}\ \bibnamefont {Rendall}},\ }\bibfield  {title} {\bibinfo {title} {The vlasov-poisson system with radiation damping},\ }in\ \href@noop {} {\emph {\bibinfo {booktitle} {Annales Henri Poincar{\'e}}}},\ Vol.~\bibinfo {volume} {2}\ (\bibinfo {organization} {Springer},\ \bibinfo {year} {2001})\ pp.\ \bibinfo {pages} {857--886}\BibitemShut {NoStop}%
\bibitem [{\citenamefont {Elskens}\ and\ \citenamefont {Kiessling}(2020)}]{Elskens}%
  \BibitemOpen
  \bibfield  {author} {\bibinfo {author} {\bibfnamefont {Y.}~\bibnamefont {Elskens}}\ and\ \bibinfo {author} {\bibfnamefont {M.-H.}\ \bibnamefont {Kiessling}},\ }\bibfield  {title} {\bibinfo {title} {Microscopic foundations of kinetic plasma theory: The relativistic vlasov--maxwell equations and their radiation-reaction-corrected generalization},\ }\href@noop {} {\bibfield  {journal} {\bibinfo  {journal} {Journal of Statistical Physics}\ }\textbf {\bibinfo {volume} {180}},\ \bibinfo {pages} {749} (\bibinfo {year} {2020})}\BibitemShut {NoStop}%
\bibitem [{\citenamefont {Noble}\ \emph {et~al.}()\citenamefont {Noble}, \citenamefont {Gratus}, \citenamefont {Burton}, \citenamefont {Ersfeld}, \citenamefont {Islam} \emph {et~al.}}]{Burton-2}%
  \BibitemOpen
  \bibfield  {author} {\bibinfo {author} {\bibfnamefont {A.}~\bibnamefont {Noble}}, \bibinfo {author} {\bibfnamefont {J.}~\bibnamefont {Gratus}}, \bibinfo {author} {\bibfnamefont {D.}~\bibnamefont {Burton}}, \bibinfo {author} {\bibfnamefont {B.}~\bibnamefont {Ersfeld}}, \bibinfo {author} {\bibfnamefont {M.~R.}\ \bibnamefont {Islam}}, \emph {et~al.},\ }\bibfield  {title} {\bibinfo {title} {Kinetic treatment of radiation reaction effects},\ }in\ \href@noop {} {\emph {\bibinfo {booktitle} {Proc. of SPIE Vol}}},\ Vol.\ \bibinfo {volume} {8079},\ pp.\ \bibinfo {pages} {80790L--1}\BibitemShut {NoStop}%
\bibitem [{\citenamefont {Berezhiani}\ \emph {et~al.}(2004)\citenamefont {Berezhiani}, \citenamefont {Hazeltine},\ and\ \citenamefont {Mahajan}}]{Mahajan1}%
  \BibitemOpen
  \bibfield  {author} {\bibinfo {author} {\bibfnamefont {V.~I.}\ \bibnamefont {Berezhiani}}, \bibinfo {author} {\bibfnamefont {R.~D.}\ \bibnamefont {Hazeltine}},\ and\ \bibinfo {author} {\bibfnamefont {S.~M.}\ \bibnamefont {Mahajan}},\ }\bibfield  {title} {\bibinfo {title} {Radiation reaction and relativistic hydrodynamics},\ }\href {https://doi.org/10.1103/PhysRevE.69.056406} {\bibfield  {journal} {\bibinfo  {journal} {Phys. Rev. E}\ }\textbf {\bibinfo {volume} {69}},\ \bibinfo {pages} {056406} (\bibinfo {year} {2004})}\BibitemShut {NoStop}%
\bibitem [{\citenamefont {Hazeltine}\ and\ \citenamefont {Mahajan}(2004)}]{Mahajan2}%
  \BibitemOpen
  \bibfield  {author} {\bibinfo {author} {\bibfnamefont {R.~D.}\ \bibnamefont {Hazeltine}}\ and\ \bibinfo {author} {\bibfnamefont {S.~M.}\ \bibnamefont {Mahajan}},\ }\bibfield  {title} {\bibinfo {title} {Closed fluid description of relativistic, magnetized plasma interacting with radiation field},\ }\href {https://doi.org/10.1103/PhysRevE.70.036404} {\bibfield  {journal} {\bibinfo  {journal} {Phys. Rev. E}\ }\textbf {\bibinfo {volume} {70}},\ \bibinfo {pages} {036404} (\bibinfo {year} {2004})}\BibitemShut {NoStop}%
\bibitem [{\citenamefont {Berezhiani}\ \emph {et~al.}(2008)\citenamefont {Berezhiani}, \citenamefont {Mahajan},\ and\ \citenamefont {Yoshida}}]{Mahajan3}%
  \BibitemOpen
  \bibfield  {author} {\bibinfo {author} {\bibfnamefont {V.~I.}\ \bibnamefont {Berezhiani}}, \bibinfo {author} {\bibfnamefont {S.~M.}\ \bibnamefont {Mahajan}},\ and\ \bibinfo {author} {\bibfnamefont {Z.}~\bibnamefont {Yoshida}},\ }\bibfield  {title} {\bibinfo {title} {Plasma acceleration and cooling by strong laser field due to the action of radiation reaction force},\ }\href {https://doi.org/10.1103/PhysRevE.78.066403} {\bibfield  {journal} {\bibinfo  {journal} {Phys. Rev. E}\ }\textbf {\bibinfo {volume} {78}},\ \bibinfo {pages} {066403} (\bibinfo {year} {2008})}\BibitemShut {NoStop}%
\bibitem [{\citenamefont {Dalakishvili}\ \emph {et~al.}(2007)\citenamefont {Dalakishvili}, \citenamefont {Rogava},\ and\ \citenamefont {Berezhiani}}]{Dalakishvili}%
  \BibitemOpen
  \bibfield  {author} {\bibinfo {author} {\bibfnamefont {G.~T.}\ \bibnamefont {Dalakishvili}}, \bibinfo {author} {\bibfnamefont {A.~D.}\ \bibnamefont {Rogava}},\ and\ \bibinfo {author} {\bibfnamefont {V.~I.}\ \bibnamefont {Berezhiani}},\ }\bibfield  {title} {\bibinfo {title} {Role of radiation reaction forces in the dynamics of centrifugally accelerated particles},\ }\href {https://doi.org/10.1103/PhysRevD.76.045003} {\bibfield  {journal} {\bibinfo  {journal} {Phys. Rev. D}\ }\textbf {\bibinfo {volume} {76}},\ \bibinfo {pages} {045003} (\bibinfo {year} {2007})}\BibitemShut {NoStop}%
\bibitem [{\citenamefont {Al-Naseri}\ and\ \citenamefont {Brodin}(2023)}]{2023RR}%
  \BibitemOpen
  \bibfield  {author} {\bibinfo {author} {\bibfnamefont {H.}~\bibnamefont {Al-Naseri}}\ and\ \bibinfo {author} {\bibfnamefont {G.}~\bibnamefont {Brodin}},\ }\bibfield  {title} {\bibinfo {title} {Radiation reaction effects in relativistic plasmas: The electrostatic limit},\ }\href@noop {} {\bibfield  {journal} {\bibinfo  {journal} {Physical Review E}\ }\textbf {\bibinfo {volume} {107}},\ \bibinfo {pages} {035203} (\bibinfo {year} {2023})}\BibitemShut {NoStop}%
\bibitem [{\citenamefont {Al-Naseri}\ and\ \citenamefont {Brodin}(2025)}]{RR2025}%
  \BibitemOpen
  \bibfield  {author} {\bibinfo {author} {\bibfnamefont {H.}~\bibnamefont {Al-Naseri}}\ and\ \bibinfo {author} {\bibfnamefont {G.}~\bibnamefont {Brodin}},\ }\bibfield  {title} {\bibinfo {title} {Probing the transition from classical to quantum radiation reaction in relativistic plasma},\ }\href@noop {} {\bibfield  {journal} {\bibinfo  {journal} {arXiv preprint arXiv:2506.22577}\ } (\bibinfo {year} {2025})}\BibitemShut {NoStop}%
\bibitem [{\citenamefont {Brodin}\ \emph {et~al.}(2023)\citenamefont {Brodin}, \citenamefont {Al-Naseri}, \citenamefont {Zamanian}, \citenamefont {Torgrimsson},\ and\ \citenamefont {Eliasson}}]{E-schwinger}%
  \BibitemOpen
  \bibfield  {author} {\bibinfo {author} {\bibfnamefont {G.}~\bibnamefont {Brodin}}, \bibinfo {author} {\bibfnamefont {H.}~\bibnamefont {Al-Naseri}}, \bibinfo {author} {\bibfnamefont {J.}~\bibnamefont {Zamanian}}, \bibinfo {author} {\bibfnamefont {G.}~\bibnamefont {Torgrimsson}},\ and\ \bibinfo {author} {\bibfnamefont {B.}~\bibnamefont {Eliasson}},\ }\bibfield  {title} {\bibinfo {title} {Plasma dynamics at the schwinger limit and beyond},\ }\href@noop {} {\bibfield  {journal} {\bibinfo  {journal} {Physical Review E}\ }\textbf {\bibinfo {volume} {107}},\ \bibinfo {pages} {035204} (\bibinfo {year} {2023})}\BibitemShut {NoStop}%
\bibitem [{\citenamefont {Al-Naseri}\ \emph {et~al.}(2021)\citenamefont {Al-Naseri}, \citenamefont {Zamanian},\ and\ \citenamefont {Brodin}}]{al2021plasma}%
  \BibitemOpen
  \bibfield  {author} {\bibinfo {author} {\bibfnamefont {H.}~\bibnamefont {Al-Naseri}}, \bibinfo {author} {\bibfnamefont {J.}~\bibnamefont {Zamanian}},\ and\ \bibinfo {author} {\bibfnamefont {G.}~\bibnamefont {Brodin}},\ }\bibfield  {title} {\bibinfo {title} {Plasma dynamics and vacuum pair creation using the dirac-heisenberg-wigner formalism},\ }\href@noop {} {\bibfield  {journal} {\bibinfo  {journal} {Physical Review E}\ }\textbf {\bibinfo {volume} {104}},\ \bibinfo {pages} {015207} (\bibinfo {year} {2021})}\BibitemShut {NoStop}%
\bibitem [{\citenamefont {Uzdensky}\ and\ \citenamefont {Rightley}(2014)}]{uzdensky2014plasma}%
  \BibitemOpen
  \bibfield  {author} {\bibinfo {author} {\bibfnamefont {D.~A.}\ \bibnamefont {Uzdensky}}\ and\ \bibinfo {author} {\bibfnamefont {S.}~\bibnamefont {Rightley}},\ }\bibfield  {title} {\bibinfo {title} {Plasma physics of extreme astrophysical environments},\ }\href@noop {} {\bibfield  {journal} {\bibinfo  {journal} {Reports on Progress in Physics}\ }\textbf {\bibinfo {volume} {77}},\ \bibinfo {pages} {036902} (\bibinfo {year} {2014})}\BibitemShut {NoStop}%
\bibitem [{\citenamefont {Melrose}\ and\ \citenamefont {Wheatland}(2016)}]{Instability}%
  \BibitemOpen
  \bibfield  {author} {\bibinfo {author} {\bibfnamefont {D.}~\bibnamefont {Melrose}}\ and\ \bibinfo {author} {\bibfnamefont {M.}~\bibnamefont {Wheatland}},\ }\bibfield  {title} {\bibinfo {title} {Is cyclotron maser emission in solar flares driven by a horseshoe distribution?},\ }\href@noop {} {\bibfield  {journal} {\bibinfo  {journal} {Solar Physics}\ }\textbf {\bibinfo {volume} {291}},\ \bibinfo {pages} {3637} (\bibinfo {year} {2016})}\BibitemShut {NoStop}%
\bibitem [{\citenamefont {Cairns}\ \emph {et~al.}(2005)\citenamefont {Cairns}, \citenamefont {Speirs}, \citenamefont {Ronald}, \citenamefont {Vorgul}, \citenamefont {Kellett}, \citenamefont {Phelps},\ and\ \citenamefont {Bingham}}]{Instability1}%
  \BibitemOpen
  \bibfield  {author} {\bibinfo {author} {\bibfnamefont {R.~A.}\ \bibnamefont {Cairns}}, \bibinfo {author} {\bibfnamefont {D.}~\bibnamefont {Speirs}}, \bibinfo {author} {\bibfnamefont {K.}~\bibnamefont {Ronald}}, \bibinfo {author} {\bibfnamefont {I.}~\bibnamefont {Vorgul}}, \bibinfo {author} {\bibfnamefont {B.}~\bibnamefont {Kellett}}, \bibinfo {author} {\bibfnamefont {A.}~\bibnamefont {Phelps}},\ and\ \bibinfo {author} {\bibfnamefont {R.}~\bibnamefont {Bingham}},\ }\bibfield  {title} {\bibinfo {title} {A cyclotron maser instability with application to space and laboratory plasmas},\ }\href@noop {} {\bibfield  {journal} {\bibinfo  {journal} {Physica Scripta}\ }\textbf {\bibinfo {volume} {2005}},\ \bibinfo {pages} {23} (\bibinfo {year} {2005})}\BibitemShut {NoStop}%
\bibitem [{\citenamefont {Bingham}\ and\ \citenamefont {Cairns}(2000)}]{Instability2}%
  \BibitemOpen
  \bibfield  {author} {\bibinfo {author} {\bibfnamefont {R.}~\bibnamefont {Bingham}}\ and\ \bibinfo {author} {\bibfnamefont {R.~A.}\ \bibnamefont {Cairns}},\ }\bibfield  {title} {\bibinfo {title} {Generation of auroral kilometric radiation by electron horseshoe distributions},\ }\href@noop {} {\bibfield  {journal} {\bibinfo  {journal} {Physics of Plasmas}\ }\textbf {\bibinfo {volume} {7}},\ \bibinfo {pages} {3089} (\bibinfo {year} {2000})}\BibitemShut {NoStop}%
\bibitem [{\citenamefont {Le~Qu{\'e}au}\ \emph {et~al.}(1984)\citenamefont {Le~Qu{\'e}au}, \citenamefont {Pellat},\ and\ \citenamefont {Roux}}]{Instability3}%
  \BibitemOpen
  \bibfield  {author} {\bibinfo {author} {\bibfnamefont {D.}~\bibnamefont {Le~Qu{\'e}au}}, \bibinfo {author} {\bibfnamefont {R.}~\bibnamefont {Pellat}},\ and\ \bibinfo {author} {\bibfnamefont {A.}~\bibnamefont {Roux}},\ }\bibfield  {title} {\bibinfo {title} {Direct generation of the auroral kilometric radiation by the maser synchrotron instability: Physical mechanism and parametric study},\ }\href@noop {} {\bibfield  {journal} {\bibinfo  {journal} {Journal of Geophysical Research: Space Physics}\ }\textbf {\bibinfo {volume} {89}},\ \bibinfo {pages} {2831} (\bibinfo {year} {1984})}\BibitemShut {NoStop}%
\end{thebibliography}%

\end{document}